\documentclass[preprint,floatfix,twocolumn,10pt]{revtex4-1} 

\usepackage{graphicx}
\usepackage{amsmath}
\usepackage{amsfonts}
\usepackage{amssymb}
\usepackage{nameref}
\usepackage{amsthm}
\usepackage{hyperref}
\usepackage{color}
\usepackage{array}

\newcommand{\bra}[1]{\ensuremath{\left\langle#1\right|}}
\newcommand{\ket}[1]{\ensuremath{\left|#1\right\rangle}}
\newcommand{\braket}[2]{\ensuremath{\left\langle#1 \vphantom{#2}\right| \left. \!\! #2 \vphantom{#1}\right\rangle}}
\newcommand{\matrixel}[3]{\ensuremath{\left\langle #1 \vphantom{#2#3} \right| #2 \left| #3 \vphantom{#1#2} \right\rangle}}

\newtheorem{theorem}{Theorem}

\graphicspath{ {./Figures/} }

\newcommand{\eqnref}[1]{Eq.~(\ref{#1})}
\newcommand{\sptsignature}[4]{\langle\langle\, #1 \,;\, #2\, ,#3 \,;\, #4 \,\rangle\rangle}

\newcommand{\onehalf}{\frac{1}{2}}

\newcommand{\timess}{\!\times\!}
\newcommand{\ztwo}{\mathbb{Z}_2}
\newcommand{\ztwotwo}{(\mathbb{Z}_2)^2}
\newcommand{\ztwothree}{(\mathbb{Z}_2)^3}
\newcommand{\ztwofour}{(\mathbb{Z}_2)^4}

\begin{document}

\title{Hierarchy of universal entanglement in 2D measurement-based quantum computation}

\author{Jacob Miller}
\email{jmilla@unm.edu}
\author{Akimasa Miyake}
\email{amiyake@unm.edu}
\affiliation{Center for Quantum Information and Control, Department of Physics and Astronomy, University of New Mexico, Albuquerque, NM 87131, USA}

\begin{abstract}
\vspace{0.4cm}
Measurement-based quantum computation (MQC) is a paradigm for studying quantum computation using many-body entanglement and single-qubit measurements. While MQC has inspired wide-ranging discoveries throughout quantum information, our understanding of the general principles underlying MQC seems to be biased by its historical reliance upon the archetypal 2D cluster state. Here, we utilize recent advances in the subject of symmetry-protected topological order (SPTO) to introduce a novel MQC resource state, whose physical and computational behavior differs fundamentally from the cluster state. We show that, in sharp contrast to the cluster state, our state enables universal quantum computation using only measurements of single-qubit Pauli $X$, $Y$, and $Z$ operators. This novel computational feature is related to the ``genuine" 2D SPTO possessed by our state, and which is absent in the cluster state. Our concrete connection between the latent computational complexity of many-body systems and macroscopic quantum orders may find applications in quantum many-body simulation for benchmarking classically intractable complexity.

\end{abstract}

\maketitle

\section{Introduction}
\label{sec:introduction}

The idea of measurement-based quantum computation (MQC), where computation is carried out solely through single-qubit measurements on a fixed many-body resource state and classical feed-forward of measurement outcomes \cite{raussendorf2001one, raussendorf2003measurement, jozsa2005introduction}, is quite surprising. This is because it highlights the origins of quantum advantage in terms of entanglement and non-commutative measurements, uniquely quantum effects without counterparts in classical mechanics. In particular, so-called universal resource states, the states that are capable of efficiently implementing universal MQC, represent a class of maximal entanglement in the classification of many-body entanglement \cite{nest2006universal}, so that the structure and complexity of their entanglement is of great interest for advancing the understanding of quantum computation. Following the canonical example of the 2D cluster state \cite{briegel2001persistent}, many other universal resource states have been found, including cluster states defined on various lattices \cite{nest2006universal}, some tensor network states \cite{verstraete2004valence, nest2006universal, gross2007novel, gross2007measurement,chen2009gapped,cai2010universal}, and model ground states in condensed matter physics such as 2D Affleck-Kennedy-Lieb-Tasaki (AKLT) states \cite{cai2010universal, miyake2011quantum, wei2011affleck, darmawan2012measurement, wei2013quantum, wei2015universal}.

Given the existence of these various known universal resource states, a natural question is whether we might be able to find any common key feature, so as to explore more their variety in fundamental structures as well as practical applications. While the earliest resource states for MQC were found in short-range correlated states described as somewhat artificial tensor network states \cite{verstraete2004valence, nest2006universal, gross2007novel, gross2007measurement,chen2009gapped,cai2010universal}, a new insight has been that a class of short-ranged entangled states structured by symmetry, endowed with so-called symmetry-protected topological order (SPTO) \cite{pollmann2010entanglement, gu2009tensor, kitaev2009periodic, ryu2010topological, chen2011classification, schuch2011classifying, pollmann2012symmetry, chen2012symmetry, chen2013symmetry}, make excellent candidate resource states systematically. Indeed, in the setting of 1D spin chains, the ground states of several SPTO phases have already been shown to possess entanglement which can be leveraged to achieve various quantum computational tasks \cite{brennen2008measurement, miyake2010quantum, bartlett2010quantum, else2012symmetryprl, else2012symmetrynjp, miller2015resource, prakash2015ground}.

Here, in adopting the concept of SPTO, we carry out such an investigation for the first time in 2D MQC, and discover a completely new kind of MQC universal resource state. Specifically, we first examine the 2D cluster state as well as a wide range of other universal resource states, and show that their 2D SPTO is trivial, of the same nature as unentangled product states. Looking more closely, we find that these previously known resource states do possess some ``weaker" SPTO, but essentially of a type closer to that of 1D spin chains. Our discovery is made possible owing to the recent progress of research into SPTO, which has revealed a hierarchy of SPTO as representing different levels of nonlocality of quantum information (see the next section for details). We then introduce our ``Union Jack" state, which in contrast possesses SPTO entirely of a 2D nature, and demonstrate that it is not only a universal resource state but additionally is ``Pauli universal," in that it can implement arbitrary quantum computation using only single-qubit measurements in the Pauli bases. As elaborated later, this feature is forbidden in the 2D cluster state on account of the Gottesman-Knill theorem \cite{gottesman1998heisenberg}, which proves the efficient classical simulability of certain quantum gates. We will conclude with the outlook that our proof of principle result about Pauli universality may be true for more general 2D SPTO resource states, which we connect to a possible deep connection between a hierarchy of SPTO in condensed matter physics and the so-called Clifford hierarchy of quantum computation.

\section{Background}
\label{sec:background}

\subsection{Measurement-based Quantum Computation and the Clifford Hierarchy}

Measurement-based quantum computation (MQC) is a means of utilizing an entangled resource state to perform computation using (generally adaptive) single-qubit measurements. Given a particular resource state, we specify our computational process by choosing a specific pattern of single-qubit measurements. Due to the probabilistic nature of measurement, different measurement outcomes will generally implement different computations. However, rather than attempting to correct for unintended measurement outcomes at every step, we can instead represent the effect of such outcomes as the product of our intended operation with a so-called byproduct operator. When these byproduct operators are sufficiently simple (e.g. Pauli operators), we can commute them through much of our computation, allowing disjoint measurements to be performed in parallel without adaptation of our measurement settings.

The canonical MQC resource state is the 2D cluster state \cite{briegel2001persistent}, which is a universal resource state, in that arbitrary quantum circuits can be simulated efficiently using an appropriate sequence of arbitrary single-spin measurements \cite{jozsa2005introduction, raussendorf2003measurement, raussendorf2001one}. The 2D cluster state is formed by preparing qubit states $\ket{+X}=\tfrac{1}{\sqrt{2}}(\ket{0}+\ket{1})$ on the vertices of a square lattice (with open boundary conditions), and applying entangling controlled-Z ($CZ$) operations, defined in the computational basis by $CZ \ket{\alpha,\beta} = (-1)^{\alpha \beta} \ket{\alpha,\beta}$, between nearest-neighbor qubits. It is described by stabilizer generators,

\begin{equation}
\label{eq:cluster_stabilizers}
S^{(i)}_C = X^{(i)} \bigotimes\limits_{j \in \text{neigh}(i)} Z^{(j)},
\end{equation}

\noindent where $\text{neigh}(i)$ is the set of nearest neighbors of site $i$. An $n$-qubit cluster state $\ket{\psi_{C}}$ is the unique state satisfying $S^{(i)}_C \ket{\psi_{C}} = \ket{\psi_{C}}$ for $i = 1,2, \ldots, n$.

The Clifford hierarchy is an ordered collection of unitary gates of increasing computational generality \cite{gottesman1999demonstrating}. The unitary gates in the $d$'th level of the Clifford hierarchy $\mathcal{C}_d$ are defined inductively, with $\mathcal{C}_1$ consisting of tensor products of Pauli operators, and $\mathcal{C}_{d+1} = \{ U |\, \forall P \!\in\! \mathcal{C}_1, U P U^\dagger \subseteq \mathcal{C}_d \}$. Each level of the Clifford hierarchy represents a greater degree of quantum-gate complexity in that, intuitively speaking, higher levels contain gates which are more ``quantum" than those in lower levels. The gates in $\mathcal{C}_2$ form a group, known as the Clifford group, which preserves the group of Pauli operators under conjugation. Exploiting this fact, the Gottesman-Knill theorem \cite{gottesman1998heisenberg} gives an efficient means of classically simulating any poly-sized circuit composed of gates in $\mathcal{C}_2$, provided that initialization and measurement occur in the single-qubit Pauli bases. By contrast, the gates in $\mathcal{C}_3$ form a universal gate set for quantum computation.

In MQC, a stronger notion of universality for resource states is Pauli universality, where the measurements used to carry out MQC are only of single-qubit Pauli operators $X$, $Y$, or $Z$. While the 2D cluster state is a universal resource state, it is formed from $CZ$ gates in $\mathcal{C}_2$ and therefore can be efficiently classically simulated when only Pauli measurements are used, making the cluster state not Pauli universal.

\subsection{Symmetry-Protected Topological Order}

Symmetry-protected topological order (SPTO) \cite{pollmann2010entanglement, gu2009tensor, kitaev2009periodic, ryu2010topological, chen2011classification, schuch2011classifying, pollmann2012symmetry, chen2012symmetry, chen2013symmetry} is a many-body phenomenon arising from many-body entanglement present in quantum states invariant under a symmetry group $G$. Given a state defined in $d$ spatial dimensions with a finite correlation length, we say that this state has nontrivial $d$-dimensional SPTO precisely when it cannot be reduced to a product state using a finite-depth quantum circuit whose gates are of constant size and commute with $G$. In this sense, nontrivial SPTO can be thought of as an indicator of persistent entanglement, protected by $G$. More generally, two $d$-dimensional states are said to be in different ($d$-dimensional) SPTO phases when they cannot be transformed into each other using such a finite-depth, symmetry-respecting quantum circuit.

\begin{figure}[t]
  \centering
  \includegraphics[width=0.48\textwidth]{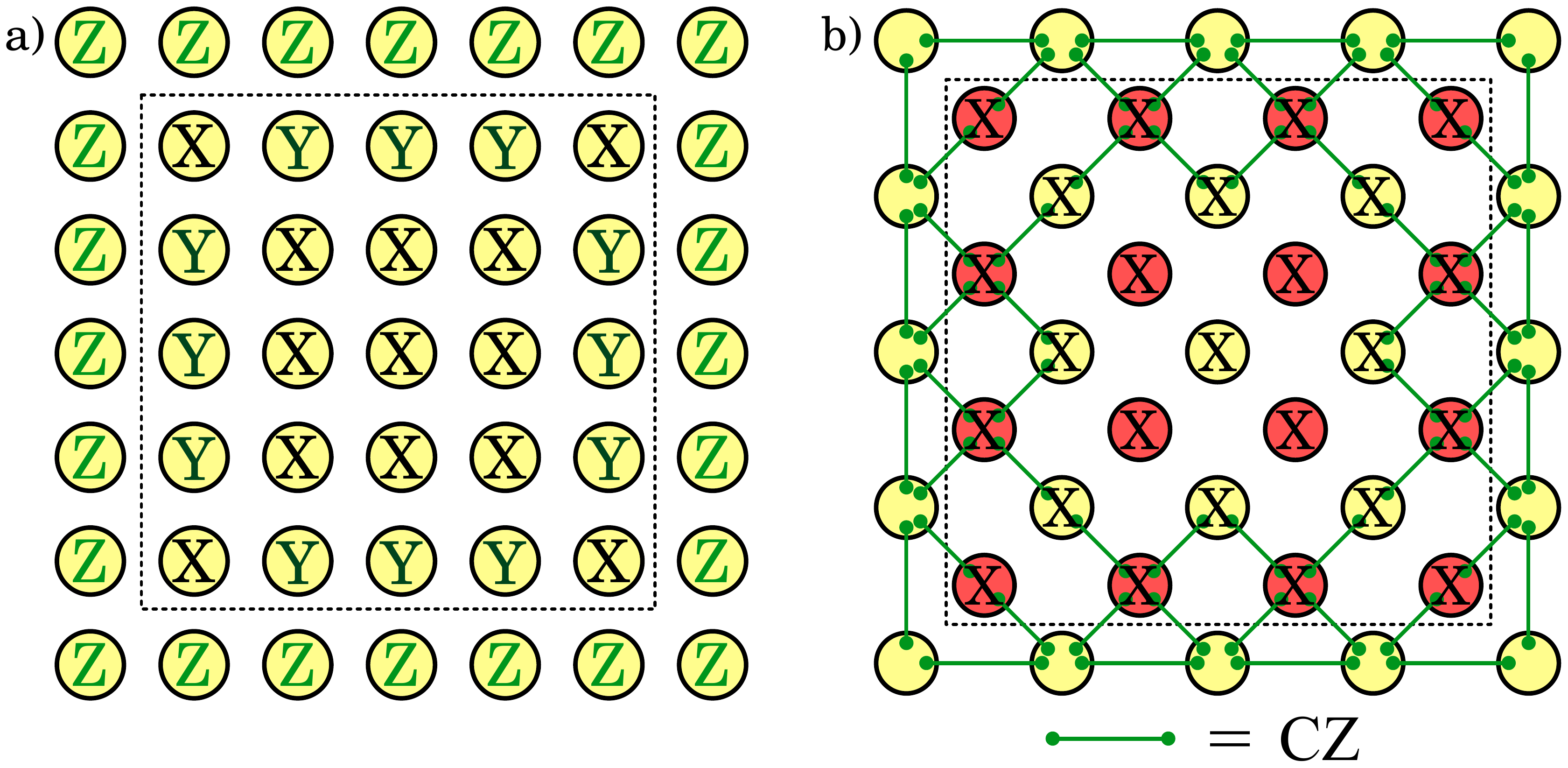}
  \caption{Manifestation of 1D and 2D SPTO in boundary symmetry operators, where $X$, $Y$, $Z$, and $CZ$ represent the application of the corresponding unitary operation. The transversal application of $X$ is only a symmetry when each state is prepared on closed boundaries. Near edges of the system, the symmetry operator must be augmented with additional boundary terms, which reflect the distinct nature of 1D vs. 2D SPTO. a) The cluster state is invariant under the application of $X$ to all sites within a region with closed boundaries. When $X$ is instead applied to a region with open boundaries (boxed area), we must add additional $Z$ (and hence $Y$, whenever $X$ and $Z$ overlap) gates near the edges to obtain a genuine symmetry. b) The Union Jack state is invariant under the application of $X$ to a region with closed boundaries. When $X$ is instead applied to a region with open boundaries (boxed area), we must add additional $CZ$ gates near the edges to obtain a genuine symmetry. The higher-dimensional SPTO manifests here as a symmetry representation which doesn't factorize into disjoint unitaries, and is built from gates at a higher level of the Clifford hierarchy.}
  \label{fig:edge_symmetry}
\end{figure}

Mathematically, $d$-dimensional SPTO phases are classified by elements of $\mathcal{H}^{d+1}(G, U(1))$, the $(d+1)$'th cohomology group of $G$, with the identity element of the group corresponding to the trivial phase of $G$-invariant product states (see Appendix~\ref{sec:spto} for an introduction to group cohomology theory). For example, when $G = \ztwo$ there is only one (trivial) 1D SPTO phase, but there are two 2D SPTO phases, one trivial and one nontrivial. Nontrivial SPTO can be detected and characterized by examining the manner in which $G$ acts on edge degrees of freedom when a state is prepared on a manifold with boundaries \cite{chen2011two, pollmann2012detection, else2014classifying, williamson2014matrix}. Nontrivial 1D SPTO manifests as a product of individual ``fractionalized" degrees of freedom on the edge, which transform under projective representations of $G$. On the other hand, nontrivial 2D SPTO manifests in the form of long-range correlated edge modes, which transform under non-separable matrix product unitary representations of $G$ \cite{chen2011two}. Concrete examples of this distinctive behavior of 1D and 2D SPTO are shown in Figure~\ref{fig:edge_symmetry}.

An important---and often neglected---fact is that states in $d$ spatial dimensions can be classified not only by a label specifying its $d$-dimensional SPTO phase, but also by other labels associated with $k$-dimensional SPTO, for $0 \leq k < d$ \cite{chen2013symmetry}. We call this collection of SPTO labels the SPTO signature of a state, denoted by $\Omega_d$ in $d$ dimensions. For $d=2$, $\Omega_2$ has the form $\Omega_2 = \sptsignature{\Theta_2}{\Theta_1^{(x)}\!\!}{\Theta_1^{(y)}}{\Theta_0}$, with $\Theta_k$ denoting a $k$-dimensional SPTO label. For general $d$, $\Omega_d$ contains $d \choose k$ $k$-dimensional SPTO labels, corresponding to the $d \choose k$ independent $k$-dimensional surfaces in $d$-dimensional space. When classifying phases, the $\Theta_k$ labels are chosen from $\mathcal{H}^{k+1}(G, U(1))$, the collection of $k$-dimensional SPTO phases for symmetry $G$. However, since we are concerned here mainly with the existence of nontrivial SPTO, we will use an abbreviated notation where $\Theta_k = 0$ or $1$ indicates trivial or nontrivial $k$-dimensional SPTO, respectively. Unlike $d$-dimensional labels, the lower-dimensional components of a state's SPTO signature can be altered by a local $G$-symmetric quantum circuit. However, these labels are unchanged by quantum circuits which respect both on-site \textit{and} lattice translational symmetries. See Appendix~\ref{sec:spto} for details about SPTO signatures.

\section{Trivial 2D SPTO of the 2D Cluster State}
\label{sec:cluster_state}

In this section, we determine the SPTO signature of the 2D cluster state, stated in Theorem~\ref{thm:cluster_state}.

\begin{figure}[t]
  \centering
  \includegraphics[width=0.48\textwidth]{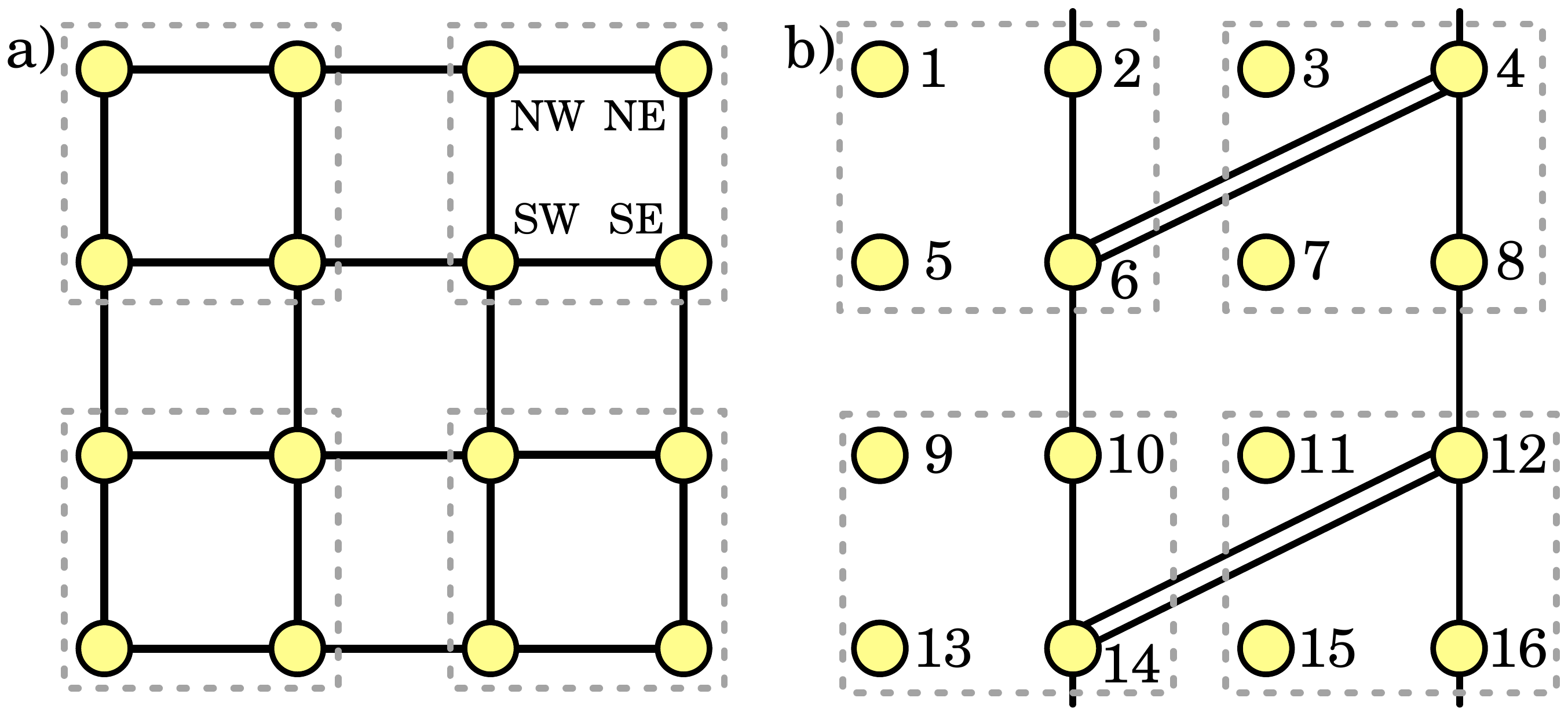}
  \caption{a) Part of the 2D cluster state on a square lattice, with $2\times2$ unit cells shown. The four generators of the $\ztwofour$ on-site symmetry are labeled. b) Part of the circuit which disentangles the 2D cluster state. Solid lines indicate a $CZ$ applied between two sites. The gate $V_E$ is shown in center, which is the product of 6 $CZ$ operations between sites $(4,8)$, $(8,12)$, $(12,14)$, $(14,10)$, $(10,6)$, and $(6,4)$. Also shown are portions of the $V_E$ gates directly above and below. Due to the ``diagonal" $CZ$'s of adjacent $V_E$'s canceling, a global tiling of these gates applies $CZ$ between all adjacent NE and SE sites. This tiling is done in two layers, so that the gates in each layer don't overlap. By applying displaced and rotated versions of these gates, we arrive at a symmetry-respecting circuit of depth 8, which disentangles the 2D cluster state to a trivial product state.}
  \label{fig:disentangler}
\end{figure}

\begin{theorem}
\label{thm:cluster_state}
    The SPTO signature of the 2D cluster state with respect to on-site $\ztwofour$ symmetry is $\Omega^{(C)}_2 = \sptsignature{0}{1}{1}{0}$, corresponding to trivial 2D SPTO and nontrivial 1D SPTO. 
\end{theorem}

The on-site $\ztwofour$ symmetry of the cluster state comes from treating a $2 \times 2$ unit cell as a single site, as shown in Figure~\ref{fig:disentangler}a. We refer to the four qubits within a unit cell by the labels NW, NE, SE, and SW. From \eqnref{eq:cluster_stabilizers}, we see that the global application of $X$ to any of these four classes of qubits preserves the cluster state stabilizers, giving the system $\ztwofour$ on-site symmetry. This is the largest on-site symmetry group of the cluster state, and its SPTO phase with respect to this group sets its SPTO phase with respect to any on-site symmetry subgroup \cite{footnote1}.

We prove the 2D part of Theorem~\ref{thm:cluster_state} by constructing a finite-depth quantum circuit, shown in Figure~\ref{fig:disentangler}b, whose gates each respect the on-site symmetry of the cluster state, but which disentangles the state to a trivial product state. Because the 2D component of a state's SPTO signature is invariant under local symmetric quantum circuits \cite{chen2013symmetry}, this suffices to prove our claim. A more careful analysis of the 2D cluster state is needed in order to prove its nontrivial 1D SPTO. In Appendix~\ref{sec:cluster_state_appendix}, we study a projected entangled pair state (PEPS) \cite{verstraete2008matrix} representation of the cluster state, which lets us characterize the transformation of its boundary under the $\ztwofour$ symmetry \cite{williamson2014matrix}. We find that individual sites along both horizontal and vertical boundaries transform under a projective representation of $\ztwofour$, giving us a ``smoking gun" indication of nontrivial 1D SPTO. This fact, demonstrated rigorously in Appendix~\ref{sec:cluster_state_appendix}, completes our proof of Theorem~\ref{thm:cluster_state}.

Importantly, a similar analysis of edge modes can be used to prove results analogous to Theorem~\ref{thm:cluster_state} for many other known universal resource states, including cluster states defined on various lattices \cite{nest2006universal} and certain 2D AKLT states \cite{cai2010universal, miyake2011quantum, wei2011affleck, darmawan2012measurement, wei2013quantum, wei2015universal}. In this sense, the impact of Theorem~\ref{thm:cluster_state} is that not just the cluster state, but in fact the majority of commonly studied universal resource states, are characterized by the absence of 2D SPTO, with at most nontrivial 1D SPTO.

\begin{figure}[t]
  \centering
  \includegraphics[width=0.48\textwidth]{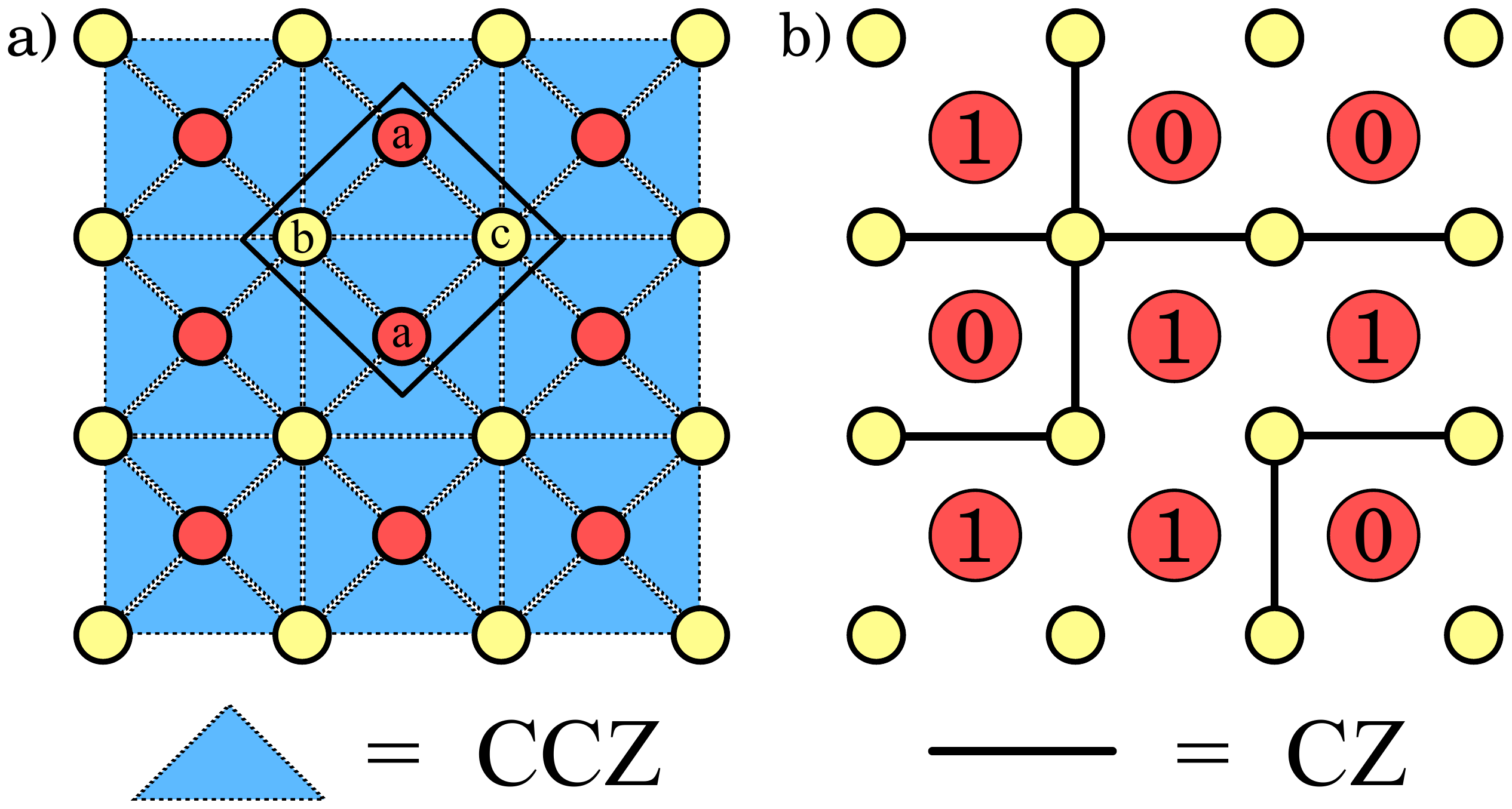}
  \caption{a) The Union Jack lattice on which our resource state is defined. Every vertex represents a qubit initialized in a $\ket{+X}$ state, and every triangular cell represents an applied 3-body unitary $CCZ$. A $2\timess2$ unit cell is shown, with respect to which our system has $\ztwothree$ symmetry generated by $X$ applied to sites a, b, or c. The $\ztwo$ symmetry of this state is a subgroup of $\ztwothree$ generated by applying $X$ to all sites. b) Measuring the control sites (red) in the computational basis collapses the remaining system into a random graph state. The edges of the graph lie on the ``domain walls" between different control site outcomes.}
  \label{fig:lattice}
\end{figure}

\section{The Resource State with Nontrivial 2D SPTO}
\label{sec:resource_state}

In this section we present a new MQC resource state that is both Pauli universal and possesses nontrivial 2D SPTO, as summarized in Theorem~\ref{thm:resource_state}. This is in contrast to the 2D cluster state, which is universal but not Pauli universal, and only possesses 1D SPTO. Our ``Union Jack" resource state is composed of qubits, each of which is located at a vertex of the Union Jack lattice shown in Figure~\ref{fig:lattice}a. It is constructed by preparing a $\ket{+X}$ state at every vertex, and then applying a 3-body doubly controlled-Z unitary operation, $CCZ$, to every triangular cell in the lattice. $CCZ$ is diagonal in the qubits' computational basis with non-zero matrix elements:

\begin{equation}
\label{eq:matrix_elements}
\matrixel{i_1 i_2 i_3}{CCZ}{i_1 i_2 i_3} = 
\begin{cases}
    -1, & \text{if } (i_1, i_2, i_3) = (1,1,1) ,\\
    +1,              & \text{otherwise} ,
\end{cases}
\end{equation}
and belongs to the 3rd level of the Clifford hierarchy $\mathcal{C}_3$. The stabilizers generated by these gates are

\begin{equation}
\label{eq:resource_stabilizers}
    S_{UJ}^{(i)} = X^{(i)} \bigotimes\limits_{(j, k) \in \text{tri}(i)} CZ^{(j, k)},
\end{equation}

\noindent where $(j, k) \in \text{tri}(i)$ refers to all pairs of sites $(j, k)$ which, together with $i$, form a triangle in the lattice of Figure~\ref{fig:lattice}a. Our resource state $\ket{\psi_{UJ}}$ is the unique state satisfying $S_{UJ}^{(i)} \ket{\psi_{UJ}} = \ket{\psi_{UJ}}$ for $i = 1,2, \ldots, n$. Note, however, that it is not a so-called stabilizer state because its stabilizer group is not contained in the $n$-qubit Pauli group.

Our resource state is an example of a ``renormalization group (RG) fixed point" state used previously to study properties of $\ztwo$ SPTO \cite{chen2013symmetry}, and consequently has $\ztwo$ symmetry. However, if we redefine a single site of our system to be a particular $2 \timess 2$ unit cell (shown in Figure~\ref{fig:lattice}a), then our system in fact has $\ztwothree$ symmetry. With respect to this latter group, our resource state can be seen as an example of a $d = 2$ decorated domain wall (DDW) state \cite{chen2014symmetry}, a method for creating systems with $d$-dimensional $\ztwo \timess G$ SPTO in terms of systems with $(d\!-\!1)$-dimensional $G$ SPTO (here $G=\ztwotwo$). We should however emphasize the importance of our state being defined on the Union Jack lattice for proving Theorem~\ref{thm:resource_state}, as the 2D state in \cite{chen2013symmetry, chen2014symmetry} is essentially defined on a triangular lattice, so that it disallows the intersection of domain walls under the procedure we use below for locally converting to a graph state \cite{hein2004multiparty, hein2005entanglement}, and thus may not be a universal resource state. On the other hand, our state is also an example of a generalization of graph states, called hypergraph states in the quantum information community \cite{rossi2013quantum, guhne2014entanglement}, although their application for MQC has not been studied previously. 

\begin{theorem}
\label{thm:resource_state}
    The Union Jack state is a Pauli universal resource state for MQC, meaning that arbitrary quantum circuits can be efficiently simulated using only measurements of single-qubit Pauli operators and feed-forward of measurement outcomes. Furthermore, its SPTO signature with respect to on-site $\ztwothree$ symmetry is $\Omega^{(UJ)}_2 = \sptsignature{1}{0}{0}{0}$, corresponding to nontrivial 2D SPTO and trivial 1D SPTO.
\end{theorem}

\begin{table}[t]
\newcommand{\colonewidth}  {2.4cm}
\newcommand{\coltwowidth}  {1.4cm}
\newcommand{\colthreewidth}{2.0cm}
\newcommand{\tworowheight} {0.72cm}
\centering
\begin{tabular}{m{\colonewidth} | m{\coltwowidth} | m{\colthreewidth}}
 & \parbox[c][\tworowheight][c]{\coltwowidth}{\centering \textbf{Cluster State}}
 & \parbox[c][\tworowheight][c]{\colthreewidth}{\centering \textbf{Union Jack State}} \\ \hline
   \parbox{\colonewidth}{\centering \textbf{SPTO}}
 & \parbox{\coltwowidth}{\centering 1D}
 & \parbox{\colthreewidth}{\centering 2D} \\ \hline
   \parbox[c][\tworowheight][c]{\colonewidth}  {\centering \textbf{Formation Circuit}}
 & \parbox[c][\tworowheight][c]{\coltwowidth}  {\centering $\mathcal{C}_2$}
 & \parbox[c][\tworowheight][c]{\colthreewidth}{\centering $\mathcal{C}_3$} \\ \hline
   \parbox[c][\tworowheight][c]{\colonewidth}  {\centering \textbf{Byproduct Operators}}
 & \parbox[c][\tworowheight][c]{\coltwowidth}  {\centering $\mathcal{C}_1$}
 & \parbox[c][\tworowheight][c]{\colthreewidth}{\centering $\mathcal{C}_2$} \\ \hline
   \parbox[c][\tworowheight][c]{\colonewidth}  {\centering \textbf{Universal Measurements}}
 & \parbox[c][\tworowheight][c]{\coltwowidth}  {\centering $\mathcal{C}_2$}
 & \parbox[c][\tworowheight][c]{\colthreewidth}{\centering $\mathcal{C}_1$}
\end{tabular}
\caption{A summary of the SPTO present in our representative resource states, the quantum circuit used to form each state, the logical byproduct operators appearing during a computation, and the single-qubit operators whose eigenbasis we need to measure to achieve universal MQC. $\mathcal{C}_d$ refers to gates chosen from the $d$'th level of the Clifford hierarchy. Higher-dimensional SPTO is associated with a higher gate complexity in the formation circuit and logical byproduct operators, and consequently requires less complexity to be added in the form of measurements. By contrast, we must perform measurements in eigenbases of gates from $\mathcal{C}_2$ in order to achieve universal MQC with the cluster state.}
\label{tab:clifford_hierarchy}
\end{table}

Note that while we phrase Theorem~\ref{thm:resource_state} in terms of our state's $\ztwothree$ SPTO, the same statement holds if we replace $\ztwothree$ by $\ztwo$. Here we demonstrate the Pauli universality of our state by efficiently simulating quantum circuits composed of Hadamard ($H$) and Toffoli ($TOFF$) gates --- a universal set of unitary gates \cite{shi2003both} --- using only measurements of single-qubit Pauli operators.
\clearpage
\onecolumngrid

\begin{figure}[t]
  \centering
  \includegraphics[width=0.86\textwidth]{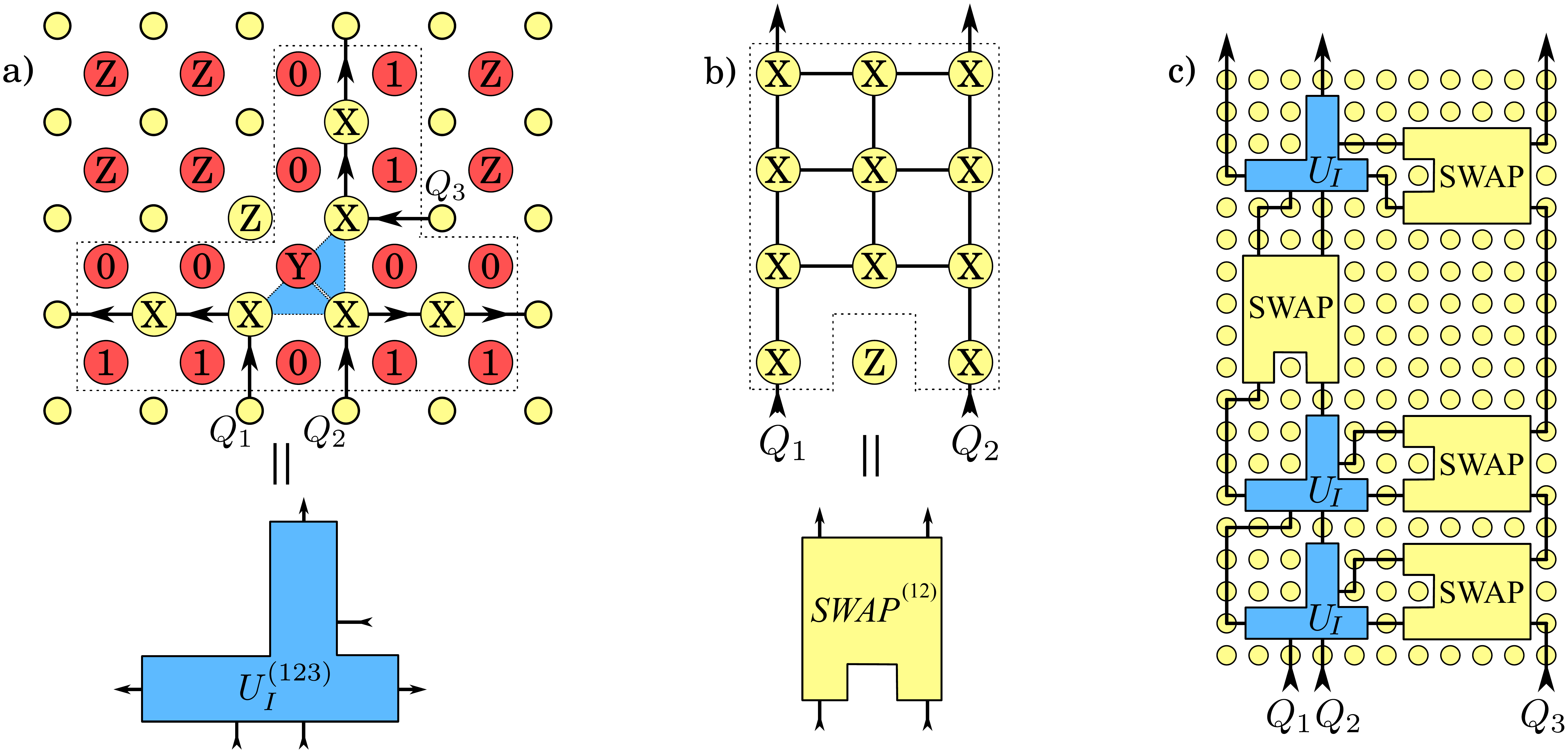}
  \caption{a) Our ``interaction gadget", which implements the non-Clifford operation $U_I$, and is formed by measuring $X$ and $Z$ on seven logical sites, $Y$ on one control site, and $Z$ on the surrounding control sites. Postselection on 13 of the latter $Z$ measurement outcomes is required in order for us to connect our gadget to the surrounding cluster region, however this only introduces a constant overhead to the number of sites measured in our protocol, as shown in Appendix~\ref{sec:protocol_details}. b) A gadget for implementing $SWAP$ within a cluster region. This allows us to implement nonplanar wire crossings, which are necessary for simulating arbitrary circuits composed of $H$ and $TOFF$ gates. c) A protocol for implementing a logical $CCZ$ gate, where solid lines indicate teleportation of logical qubits. The majority of the sites involved have been converted to an extended cluster region, with the exception of the sites used to construct interaction gadgets. Our diagram only reflects the topology of the relevant logical connections, whereas a realistic implementation would involve a detailed measurement pattern to perform teleportation throughout the cluster region, as well as a significantly greater distance between neighboring gadgets. More explicit details of our protocol can be found in Appendix~\ref{sec:protocol_details}.}
  \label{fig:protocol_details}
\end{figure}

\twocolumngrid

Our means of simulating these circuits using the Union Jack state are divided into two parts. We first show that portions of our state can be converted into ``cluster regions", regions which are locally identical to the 2D cluster state. These cluster regions are used to prepare and readout qubit states, teleport these states over arbitrarily long distances, and apply Clifford gates (which include $H$ gates) to them, all with the use of only Pauli measurements. We then demonstrate that we can implement $CCZ$ using certain ``interaction gadgets", which are prepared using Pauli measurements. Since we can implement both $H$ and $CCZ$ gates, and because $TOFF$ and $CCZ$ are related by $TOFF^{(123)} = H^{(3)} CCZ^{(123)} H^{(3)}$, the combination of cluster regions and interaction gadgets lets us implement $H$ and $TOFF$ gates, and therefore arbitrary quantum circuits.

\begin{figure}[t]
  \centering
  \hbox{\hspace{-0.3cm}
  \includegraphics[width=0.55\textwidth]{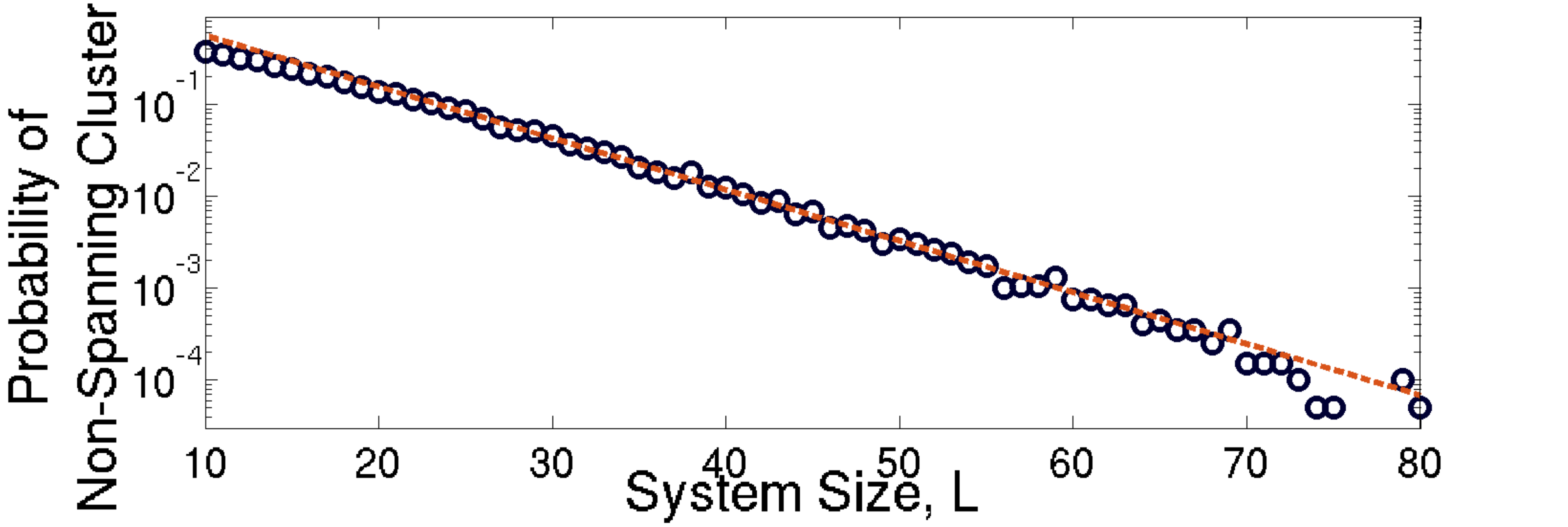}}
  \caption{A simulation of our percolation problem with increasing linear size, $L$. The exponential decay of the non-spanning probability is characteristic of the percolation supercritical phase, demonstrating that portions of our Union Jack state can be locally reduced to a 2D cluster state with arbitrarily high probability. These cluster regions are used to perform Clifford operations upon our computational qubits, as well as to shuttle these qubits between spatially separated interaction gadgets, which can be connected together to produce logical $CCZ$ gates.}
  \label{fig:vary_size}
\end{figure}

Our technique for creating cluster regions within the Union Jack state is to induce a symmetry-breaking phase transition from 2D to 1D SPTO. This involves first performing a computational basis measurement of all the Union Jack control qubits, shown in Figure~\ref{fig:lattice}b. This symmetry-breaking measurement forces the remaining part of our system, which lives on a regular square lattice, into a random graph state whose edges (associated with nontrivial 1D SPTO) appear along the domain walls in our measurement outcomes. In particular, we obtain an edge ($CZ$ gate) in our graph whenever two adjacent measurement outcomes differ, and no edge whenever they agree. We can then use the exact same protocol as in \cite{browne2008phase} to reduce this random graph state to a state which is locally identical to the regular 2D cluster state. This protocol succeeds with a probability that converges exponentially fast to either 0 or 1 in the limit of large cluster regions, depending on whether our random graph state percolates and has a macroscopic spanning cluster of connected vertices. We perform numerical simulations of this percolation problem for different system sizes, and conclude (see Figure~\ref{fig:vary_size}) that our Union Jack system is in a supercritical percolation phase and thus can be used to efficiently prepare connected cluster regions.

Our technique for preparing interaction gadgets involves taking a small area of the Union Jack state and applying an appropriate pattern of Pauli measurements to it (see Figure~\ref{fig:protocol_details}a). When embedded within a cluster region, these gadgets implement a three-body non-Clifford logical gate $U_I$, defined by $U_I^{(123)} = CCZ^{(123)} \sqrt{CZ}^{(12)} \sqrt{CZ}^{(23)}$, where $\sqrt{CZ}$ acts as $\sqrt{CZ} \ket{\alpha,\beta} = (i)^{\alpha \beta} \ket{\alpha,\beta}$. Using $U_I$, we can obtain $CCZ$ by applying $U_I$ three times to the same triple of qubits, but with the qubits cyclically permuted each time. This permutation involves crossing adjacent wires, something which is forbidden in a strictly planar graph structure, but we can simulate a nonplanar wire crossing using a $SWAP$ operation within our cluster regions (see Figure~\ref{fig:protocol_details}b). The identity $U_I^{(123)} U_I^{(231)} U_I^{(312)} =\nobreak CCZ^{(123)} CZ^{(12)} CZ^{(13)} CZ^{(23)}$ shows that this gives the desired operation of $CCZ$, up to byproduct $CZ$ gates. These byproduct gates, as well as other byproduct Clifford gates which appear in our protocol, are adaptively eliminated within cluster regions by applying the appropriate inverse Clifford operations. This adaptive cancellation of byproduct operators is generally necessary before the application of subsequent $H$ or $CCZ$ logical gates, since attempting to commute them through these gates would lead to a byproduct group which doesn't close at any level of the Clifford hierarchy. Additional information about our protocol for establishing Pauli universality of the Union Jack state is given in Appendix~\ref{sec:protocol_details}.

Looking at the proof just given, we see that the disparity between the universality of the cluster state and the Pauli universality of the Union Jack state arises from the difference in gates implementable by each state under Pauli measurements, $\mathcal{C}_2$ for the former and $\mathcal{C}_3$ for the latter. Some insight can be gained by comparing this computational difference to the fact that the cluster state and Union Jack state respectively possess 1D and 2D SPTO, as summarized in Table~\ref{tab:clifford_hierarchy}. Generalizing from these examples, we might expect this correspondence between SPTO and the Clifford hierarchy to extend to a wider class of SPTO states, providing a general link between types of SPTO and degrees of quantum gate complexity. Such a correspondence was demonstrated in \cite{yoshida2015topological, yoshida2015gapped} for topological quantum error-correcting codes, but proving this in the setting of MQC would give a means of directly associating computational characteristics to SPTO states, without the need for an auxiliary higher-dimensional topologically ordered system.

\vspace{0.3cm}
\section{Discussion}
\label{sec:outlook}

Although pertaining most immediately to MQC, our main results can be fruitfully interpreted as general statements about the interplay of two intrinsically quantum ingredients, entanglement and measurement, which play a leading role in quantum information science. Our Theorem~\ref{thm:cluster_state} demonstrates that previously studied resource states, despite differing in their microscopic details, possess identical forms of macroscopic entanglement, namely 1D SPTO. While such entanglement is sufficient for universal quantum computation using arbitrary single-qubit measurements, our Theorem~\ref{thm:resource_state} demonstrates that the use of more complex forms of entanglement, namely 2D SPTO, lets us achieve the same results using simpler Pauli measurements. As argued in the previous section, we expect that this tradeoff between entanglement and measurement is not only true of more general quantum systems, but in fact is evidence of a deep connection between the hierarchies of SPTO and the Clifford hierarchy of quantum computation. Such a connection between the computational complexity of many-body systems and their emergent macroscopic behavior would give a means of converting canonical condensed matter tools, such as order parameters, into interesting indicators of computational behavior, as was done with 1D spin chains in \cite{miller2015resource}. The natural connection we demonstrate between the computational complexity of many-body systems and emergent macroscopic order may find applications for better understanding the emergence of classically intractable complexity within quantum many-body simulation \cite{cirac2012goals, georgescu2014quantum}.

\section{Acknowledgements}
This work was supported in part by National Science Foundation grants PHY-1212445, PHY-1314955, PHY-1521016, and PHY-1620651.

\appendix
\vspace{1.2cm}

In Appendix~\ref{sec:spto} we provide a more complete discussion of SPTO, including an overview of the relevant concepts from group cohomology theory. In Appendix~\ref{sec:other_models} we demonstrate the precise manner in which the 2D cluster state and the Union Jack state are examples of SPTO fixed-point states originally introduced in \cite{chen2013symmetry}. Finally, in Appendices~\ref{sec:cluster_state_appendix} and \ref{sec:protocol_details} we give the full proofs of our Theorems~\ref{thm:cluster_state} and \ref{thm:resource_state}.

\section{Symmetry-Protected Topological Order}
\label{sec:spto}

We give here a more complete discussion of SPTO, and in particular the possible SPTO signatures that are allowed for an arbitrary 2D state. We restrict our discussion to systems with an on-site symmetry $G$, and ignore SPTO arising from global symmetries, such as time reversal, spatial inversion, or lattice point group symmetries. However, we do consider the effect of lattice translational symmetries, since this symmetry is necessary for lower-dimensional portions of our SPTO signature to be well-defined. After having given this general discussion of SPTO, we state the classification of several SPTO phases in 2D and 1D which are relevant for our purposes.

The classification of SPTO phases is closely tied to group cohomology theory, so we first give a brief introduction to some of the concepts from that field. Given a symmetry group $G$, we can construct $n$-cochains $\omega_n$, which are functions from the direct product of $n$ copies of $G$ to the group of complex phases, $U(1) = \{ \alpha \in \mathbb{C}\,|\,\alpha \alpha^* = 1 \}$. The collection of $n$-cochains form an abelian group $\mathcal{C}^n(G, U(1))$ under pointwise multiplication, with the product of cochains $\omega_n$ and $\omega'_n$ given by a cochain $\omega_n \omega'_n$, where $(\omega_n \omega'_n)(g_1,\ldots,g_n) = \omega_n(g_1,\ldots,g_n) \omega'_n(g_1,\ldots,g_n)$. The identity element in $\mathcal{C}^n(G, U(1))$ is the trivial $n$-cochain, $\omega_n^0(g_1,\ldots,g_n) = 1$. We define an operation called the coboundary operator, $d_n: \mathcal{C}^n(G, U(1)) \to \mathcal{C}^{n+1}(G, U(1))$, by

\begin{multline}
\label{eq:coboundary_operator}
(d_n \omega_n)(g_1,\ldots,g_{n+1}) = \\
    \omega_n(g_2,\ldots,g_{n+1}) \omega_n^{(-1)^{n+1}}(g_1,\ldots,g_n) \\
    \prod_{k=1}^n \omega_n^{(-1)^k}(g_1,\ldots,g_{k-1},g_k g_{k+1},g_{k+2},\ldots,g_{n+1}) .
\end{multline}

A special role is played by the $n$-cocycles and $n$-coboundaries, which form subgroups of $\mathcal{C}^n(G, U(1))$ denoted by $\mathcal{Z}^n(G, U(1))$ and $\mathcal{B}^n(G, U(1))$, respectively. An $n$-cochain is an $n$-cocycle (resp. $n$-coboundary) if it lies in the kernel of $d_n$ (resp. the image of $d_{n-1}$). More explicitly, $\mathcal{Z}^n(G, U(1)) = \{\,\omega_n\,|\,d_n \omega_n = \omega_{n+1}^0 \}$ and $\mathcal{B}^n(G, U(1)) = \{\,\omega_n\,|\,\exists \omega_{n-1} \text{ s.t. } d_n \omega_{n-1} = \omega_n \}$. One can show that the composite of coboundary operators $d_n$ and $d_{n+1}$ is trivial, in that it sends every $n$-cochain to the identity $(n+2)$-cochain. This implies that every $n$-coboundary is an $n$-cocycle, so that $\mathcal{B}^n(G, U(1)) \subseteq \mathcal{Z}^n(G, U(1))$.

We define the $n$'th cohomology group of $G$, $\mathcal{H}^n(G, U(1))$, to be the (abelian group) quotient of $\mathcal{Z}^n(G, U(1))$ with respect to $\mathcal{B}^n(G, U(1))$, $\mathcal{H}^n(G, U(1)) = \mathcal{Z}^n(G, U(1)) / \mathcal{B}^n(G, U(1))$. Equivalently, this is the group of equivalence classes of $n$-cocycles, $\mathcal{H}^n(G, U(1)) = \{\,[\omega_n]\,|\,\omega_n \in \mathcal{Z}^n(G, U(1)) \}$, under the equivalence relation $[\omega_n] = [\omega'_n] \Leftrightarrow \omega_n = \omega'_n \omega''_n$, where $\omega''_n$ is an arbitrary $n$-coboundary. For $\omega_n \in \mathcal{Z}^n(G, U(1))$, we will call $[\omega_n] \in \mathcal{H}^n(G, U(1))$ the cohomology class associated to $\omega_n$.

The relevance of this discussion for our purposes is that SPTO phases of $G$-invariant many-body systems living in $d$-dimensional space are classified by elements of the $(d+1)$'th cohomology group. In particular, it was shown in \cite{chen2013symmetry} that given any two distinct cohomology classes in $\mathcal{H}^{(d+1)}(G, U(1))$, we can construct $d$-dimensional ``fixed point" systems labeled by the cohomology classes which belong to different SPTO phases. This construction is discussed in more detail in Section~\ref{sec:other_models}.

An important point is that systems with both on-site $G$ symmetry and translational symmetry admit a richer classification of SPTO phases \cite{chen2013symmetry}. In particular, while the SPTO phase of a system without translational symmetry can be uniquely classified by a single cohomology class, with additional translational symmetry in place, the SPTO phase is classified by a full SPTO signature $\Omega_d$, which consists of an ordered list of different cohomology classes. For systems in 2D, this signature is of the form $\Omega_2 = \sptsignature{[\omega_3]}{[\omega_2^{(x)}]}{[\omega_2^{(y)}]}{[\omega_1]}$, with $[\omega_3] \in \mathcal{H}^3(G, U(1))$, $[\omega_2^{(x)}], [\omega_2^{(y)}] \in \mathcal{H}^2(G, U(1))$, and $[\omega_1] \in \mathcal{H}^1(G, U(1))$. We refer to these respectively as the 2D, 1D, and 0D portions of $\Omega_2$. For SPTO systems in $d$ physical dimensions, there will generally be ${d \choose k}$ components to the $k$-dimensional sector of the SPTO signature, corresponding to the number of independent $k$-dimensional surfaces in $d$-dimensional space. Due to our present focus on only whether or not a system possesses SPTO, we often use an abbreviated means of writing the components of an SPTO signature, wherein a phase label is written as 0 if it corresponds to the trivial phase, and as 1 if it corresponds to any nontrivial phase.

We now introduce a few examples of concrete SPTO phases in 2D and 1D associated with various symmetry groups. Since there is always a trivial phase for every symmetry group and dimension, we will often neglect to mention these phases.

For $G = \ztwo$, we have no nontrivial phases in 1D, and one nontrivial phase in 2D. Our Union Jack state lives in this nontrivial 2D $\ztwo$ phase when its symmetry group is taken to be $\ztwo$.

For $G = D_2 \simeq \ztwotwo$, we have one nontrivial phase in 1D (known as the $D_2$ Haldane phase), and 7 nontrivial phases in 2D. $D_2$ is the smallest symmetry group which is capable of manifesting SPTO in 1D.

For $G = \ztwothree$, we have 7 nontrivial phases in 1D and 127 nontrivial phases in 2D. Using a known decomposition of 2D abelian SPTO phases (those with $G$ abelian), we can structure the 2D $\ztwothree$ phases as $\mathcal{H}^3(\ztwothree, U(1)) \simeq \ztwothree \times \ztwothree \times \ztwo$ \cite{zaletel2014detecting}. The first (resp. second) $\ztwothree$ factor encodes the ``type I" (resp. ``type II") phases, those whose nontrivial SPTO arises from only one (resp., from pairs) of the $\ztwo$ components in $\ztwothree$. The last $\ztwo$ in the decomposition of $\mathcal{H}^3(\ztwothree, U(1))$ is the unique ``type III" component of the phase, which is due to a nontrivial combination of all three $\ztwo$ components in $\ztwothree$. Our Union Jack state with $\ztwothree$ symmetry belongs to the phase $(0, 0, 1)$, meaning the unique phase with trivial type I and II SPTO, and nontrivial type III SPTO.

\section{The Union Jack and Cluster States as SPTO Fixed Point States}
\label{sec:other_models}

In this Section, we demonstrate how both the Union Jack and 2D cluster states are examples of the construction of \cite{chen2013symmetry} for constructing special RG fixed point states with nontrivial SPTO from nontrivial cocycles of a symmetry group $G$. We show how our Union Jack state belongs to this class of states both for $G = \ztwo$ and for $G = \ztwothree$, and how the 2D cluster state belongs to this class of states.

The construction of \cite{chen2013symmetry} gives a means of taking $d$-dimensional SPTO signatures, along with a representative $(k+1)$-cocycle for each $k$-dimensional component of the signature, and constructing a $d$-dimensional state with that SPTO signature. For our purposes, we will focus on $d = 2$, for which the 2D, 1D, and/or 0D labels are allowed to be nontrivial. We will restrict first to the case of trivial lower-dimensional SPTO (the case considered almost exclusively in \cite{chen2013symmetry}), and later explain how these lower-dimensional labels can be made nontrivial.

To construct a 2D state from a chosen group $G$ and 3-cocycle $\omega_3$, we first choose a triangulated 2D lattice on which our state will live, and assign a Hilbert space $H_G$ to every lattice vertex. $H_G$ has dimension $|G|$, the order of $G$, and is spanned by an orthonormal basis labeled by the elements of $G$, $\{ \ket{g} \}_{g \in G}$. $G$ acts on $H_G$ as the regular representation $u_G$, with $u_g \ket{h} = \ket{gh}$ for every $g,h \in G$. We first initialize every $H_G$ in the unique invariant state $\ket{\phi_G} = (1 / \sqrt{|G|}) \sum_{g \in G} \ket{g}$, which gives a symmetric global product state with trivial SPTO. We then apply to this system a collection of 3-body unitary gates, each formed from our chosen 3-cocycle, which generates the nontrivial 2D SPTO. The 3-body unitary $\hat{\omega}_3$ generated from a 3-cocycle $\omega_3$ is diagonal in the $G$-basis, and has non-zero matrix elements of

\begin{equation}
\label{eq:general_cocycle_matrix_elements}
    \matrixel{g h f}{\hat{\omega}_3}{g h f} = \omega_3(g, g^{-1}h, h^{-1}f).
\end{equation}

\noindent Our desired state is obtained by applying $\hat{\omega}_3$ or its inverse to the vertices around every triangular cell in our chosen lattice. Whether we apply $\hat{\omega}_3$ or $\hat{\omega}_3^\dagger$ to a particular triangular cell, as well as how we match up the 3 indices in \eqnref{eq:general_cocycle_matrix_elements} with the three sites around that cell, depend on a certain ordering of lattice vertices. While the full details are given in \cite{chen2013symmetry}, if we restrict to 3-colorable lattices we can always choose each of the three indices to be matched up with a different vertex color in a fixed manner.

Choosing $G = \ztwo \simeq \{0, 1\}$, this construction outputs qubit states, with $\ket{\phi_G} = \ket{+X}$. To produce our Union Jack state, we work with the Union Jack lattice, and choose our 3-cocycle to be

\begin{equation}
\label{eq:z2_cocycle}
\omega_3(g,h,f) = 
\begin{cases}
    -1, & \text{if } (g, h, f) = (1,1,1) \\
    +1,              & \text{otherwise}.
\end{cases}
\end{equation}

\noindent Although this 3-cocycle produces a unitary $\hat{\omega}_3$ which is distinct from $CCZ$, the global state it produces is nonetheless the same. This can be seen from the relation $\hat{\omega}_3^{(123)} = CCZ^{(123)} CZ^{(13)}$, which allows us to show that the transversal application of $\hat{\omega}_3$ to qubits in any 3-colorable lattice with closed (nonexistent) boundary yields the same global unitary as the transversal application of $CCZ$. This proves that the Union Jack state is a $\ztwo$ SPTO fixed point state, associated with the cocycle of \eqnref{eq:z2_cocycle}. Because this 3-cocycle belongs to the unique nontrivial cohomology class in $\mathcal{H}^3(\ztwo, U(1))$, our Union Jack state consequently has nontrivial 2D SPTO.

Showing that our Union Jack state is isomorphic to a $\ztwothree$ SPTO fixed point state is less obvious, since the lattice vertices of such states aren't associated with qubits, but rather with 8-dimensional qudits. We can get around this difficulty by first treating each of the $\ztwo$ factors in $\ztwothree$ as a separate qubit system, and imagining these three factors to be stacked vertically in three layers at each lattice site. Note that this stacking is merely a convenient means of visualizing the separate qubit factors in $\ztwothree$, while our lattice remains a genuine 2D lattice. In this case, the state we initialize each site in is $\ket{\phi_G} = \ket{+X}^{\otimes 3}$, a tensor product of one $\ket{+X}$ state on each layer. If we write a generic element $g \in \ztwothree$ as $g = (g_1, g_2, g_3)$, where each $g_i \in \ztwo$ is associated with the $i$'th layer, then we can choose the following 3-cocycle

\begin{equation}
\label{eq:z23_cocycle}
\omega_3' (g,h,f) = 
\begin{cases}
    -1, & \text{if } (g_1, h_2, f_3) = (1,1,1) \\
    +1,              & \text{otherwise},
\end{cases}
\end{equation}

\noindent where addition is modulo 2. Using the relation \eqnref{eq:general_cocycle_matrix_elements}, we can show that $\hat{\omega}_3'$ equals a $CCZ$ gate on the qubits indexed by $g_1$, $h_2$, and $f_3$, along with other terms which cancel when $\hat{\omega}_3'$ is applied globally. In other words, $\hat{\omega}_3'$ ends up having a nontrivial action only on the qubits on the first layer of the first site acted on, the second layer of the second site, and the third layer of the third site. If we apply $\hat{\omega}_3'$ transversally to all triangular cells on a 3-colorable lattice, then at each site only one of the three layers is acted on nontrivially, with the other two layers remaining unchanged. Thus, using $\hat{\omega}_3'$ to construct a $\ztwothree$ SPTO fixed point state defined on a Union Jack lattice with $n$ vertices yields a state which is a tensor product of our Union Jack state on $n$ qubits, with $\ket{+X}$ on the remaining $2n$ qubits. This proves that, up to addition/removal of ancilla $\ket{+X}$ states, the Union Jack state is a $\ztwothree$ SPTO fixed point state, associated with the cocycle of \eqnref{eq:z23_cocycle}. This cocycle belongs to the nontrivial $\ztwothree$ cohomology class described at the end of Section~\ref{sec:spto}, which consequently specifies the nontrivial $\ztwothree$ SPTO phase our Union Jack state belongs to.

As the 2D cluster state only possesses lower-dimensional SPTO, we must use an extended version of the previous construction to obtain the cluster state as an SPTO fixed point state. In \cite{chen2013symmetry} it is shown that to generate 2D fixed point states with 1D SPTO, we can use a construction almost identical to that given above, but instead of starting with a 3-cocycle $\omega_3$ and converting it into a 3-body gate $\hat{\omega}_3$, we start with a 2-cocycle $\omega_2$ and convert it into a 2-body gate $\omega_2$, which has non-zero matrix elements of

\begin{equation}
\label{eq:general_2_cocycle_matrix_elements}
    \matrixel{g h}{\hat{\omega}_2}{g h} = \omega_2(g, g^{-1}h).
\end{equation}

\noindent $\hat{\omega}_2$ is then applied to all edges of a chosen 2D lattice, on which one copy of $\ket{\phi_G}$ has been prepared at every vertex. To generate the 2D cluster state in this manner, we can choose $G = \ztwotwo$ and use a similar decomposition of the local Hilbert space into two qubits, stacked vertically in two layers. We then utilize the 2-cocycle

\begin{equation}
\label{eq:z22_cocycle}
\omega_2 (g,h) = 
\begin{cases}
    -1, & \text{if } (g_1, h_2) = (1,1) \\
    +1,              & \text{otherwise},
\end{cases}
\end{equation}

\noindent where $g_i, h_i \in \ztwo$ is associated with the $i$'th component of $g, h \in \ztwotwo$. This 2-cocycle produces a 2-body unitary $\hat{\omega}_2$ which upon global application is equivalent to a $CZ$ gate on the qubits indexed by $g_1$ and $h_2$, and an identity gate on the rest of the qubits. In close analogy to how the Union Jack state was shown above to be a $\ztwothree$ SPTO fixed point state, we can work with the 2-colorable square lattice and show that the transversal application of $\hat{\omega}_2$ to all edges of the lattice yields a state which is a tensor product of the 2D cluster state on $n$ qubits, with $\ket{+X}$ on the remaining $n$ qubits. This proves that, up to addition/removal of ancilla $\ket{+X}$ states, the cluster state is a $\ztwotwo$ SPTO fixed point state.

Finally, we note that some care is required regarding the symmetry group of the 2D cluster state. The construction we just outlined outputs the cluster state as an SPTO fixed point state with $\ztwotwo$ symmetry, similar to how the 1D cluster state is most naturally seen as possessing nontrivial SPTO associated with $\ztwotwo$ symmetry. However, as seen from \eqnref{eq:virtual_representation}, if we choose any particular $\ztwotwo$ subgroup of the full $\ztwofour$ on-site symmetry, we obtain a virtual representation of our symmetry which is non-projective in at least one direction. This leads to an SPTO signature which is either $\sptsignature{0}{0}{1}{0}$ or $\sptsignature{0}{1}{0}{0}$, rather than the SPTO signature of $\sptsignature{0}{1}{1}{0}$ which appears in Theorem~\ref{thm:cluster_state}. We interpret this fact as an indicator that for states with lower-dimensional SPTO, we must take care in choosing the symmetry group we use to arrive at an SPTO signature.

\section{SPTO Signature of the 2D Cluster State}
\label{sec:cluster_state_appendix}

We present here a full demonstration that the SPTO signature of the 2D cluster state is $\Omega_2^{(C)} = \sptsignature{0}{1}{1}{0}$, as stated in Theorem~\ref{thm:cluster_state}. To do this, we need to determine the various cohomology classes corresponding to different components of the cluster state's signature. One known way \cite{williamson2014matrix} of doing this is by working with a projected entangled pair state (PEPS) description of the cluster state, and examining the behavior of the representation of its on-site symmetry group $\ztwofour$ along the boundary.

\begin{figure}[t]
  \centering
  \hbox{\hspace{0cm}
  \includegraphics[width=0.5\textwidth]{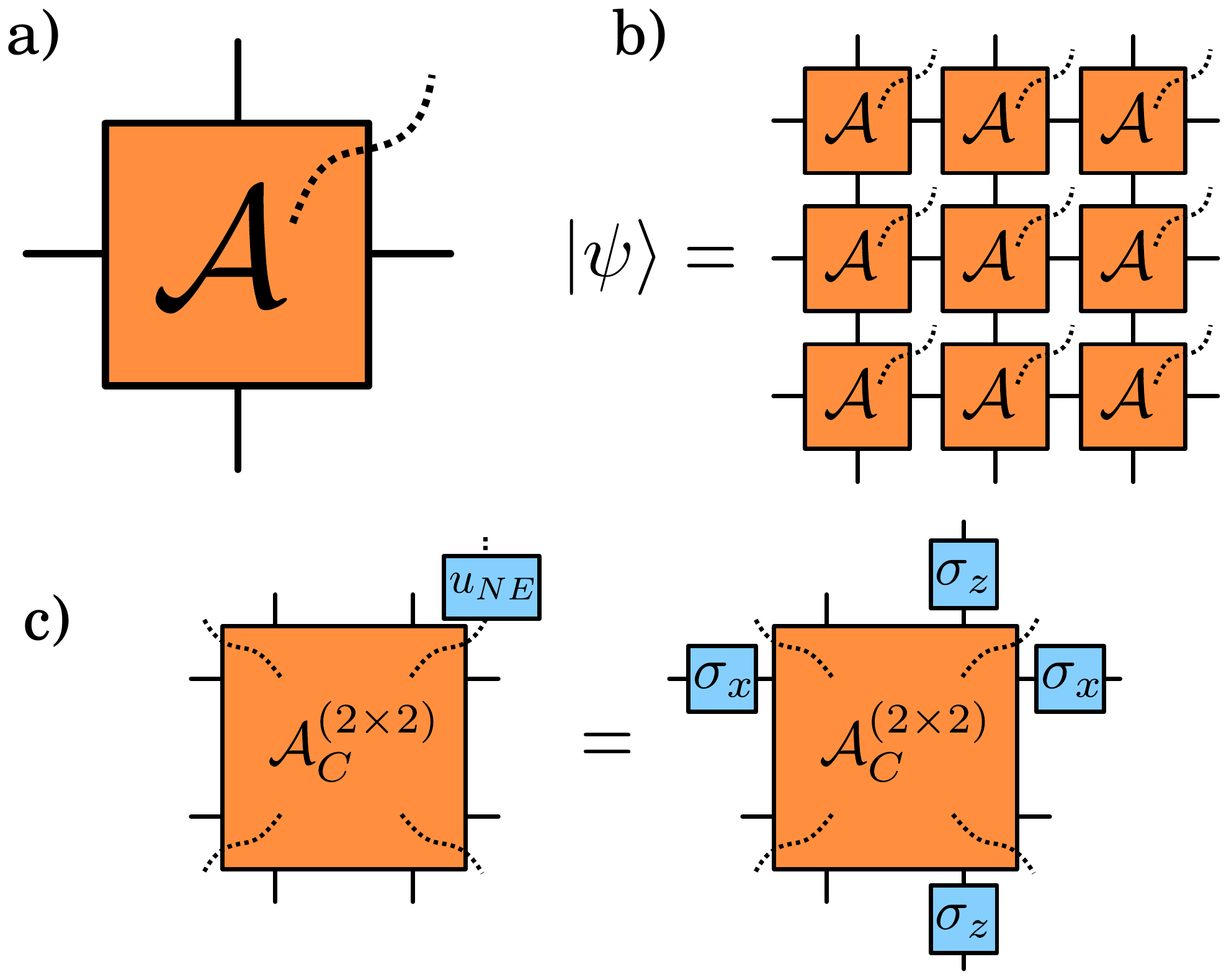}}
  \caption{a) A single PEPS tensor for a square lattice. The dotted line represents our physical system, which corresponds to a single site of our lattice, and the four solid edges represent the virtual space. b) After assigning a PEPS tensor to every site of our lattice, we obtain a physical state by taking the ``tensor trace" of all tensors. This involves contracting every pair of adjacent virtual indices using a maximally entangled state $\ket{\phi_0} = \sum_{i=1}^{D_{v}} \ket{i, i}$, with $D_{v}$ the virtual space dimension. On a lattice with no boundary, this will contract out all of the virtual spaces, leaving only our physical many-body state $\ket{\psi}$. c) An example of the physical/virtual symmetry correspondence given in \eqnref{eq:virtual_symmetry} for the 2D cluster state. Our PEPS tensor is defined relative to a $2\timess2$ physical unit cell, with a four-qubit physical space and two-qubit virtual spaces. Different generators of $\ztwofour$ will produce different combinations of $X$ and $Z$ on the virtual space, whose noncommutativity demonstrates the nontrivial 1D SPTO of the 2D cluster state.}
  \label{fig:peps}
\end{figure}

Restricting to states which live on a square lattice, a PEPS representation consists of a rank-5 tensor, $\mathcal{A} \in H_{p} \otimes (H_{v}^*)^{\otimes 4}$, where $H_{p}$ and $H_{v}$ are referred to as the physical and virtual Hilbert spaces, and where $H^*$ denotes the Hilbert space dual to $H$. $\mathcal{A}$ can also be interpreted as a map $\mathcal{A}: H_{p}^* \to (H_{v}^*)^{\otimes 4}$. We associate one copy of $\mathcal{A}$ to each site of our lattice, with $H_{p}$ corresponding to the Hilbert space of that site, and the four $H_{v}^*$'s being used to represent correlations between our site and each of the four nearest-neighbor sites. The dimension of $H_{v}$, $D_{v}$, is the bond dimension of our PEPS representation, and can be thought of as a measure of entanglement in the system. The condition for $\mathcal{A}$ to be a PEPS representation of a many-body state $\ket{\psi}$ is that the ``tensor trace" of the $\mathcal{A}$'s at every site, formed by contracting every pair of adjacent $H_{v}^*$'s using maximally entangled states $\ket{\phi_0} = \sum_{i=1}^{D_{v}} \ket{i, i}$, yields $\ket{\psi}$. This condition is depicted in Figure~\ref{fig:peps}b.

Given a PEPS representation $\mathcal{A}$ of our many-body state $\ket{\psi}$, the condition for $\ket{\psi}$ to be invariant under our on-site symmetry $G$, whose physical representation is $u_G = \{\,u_g\,|\,g \in G \}$, is that there exists a virtual representation of $G$, $\mathcal{U}_G$, such that

\begin{equation}
\label{eq:virtual_symmetry}
\mathcal{A} \,u_G = e^{i \theta_G}\, \mathcal{U}_G \mathcal{A}.
\end{equation}

In other words, when $\mathcal{A}$ is seen as a map from the physical to the virtual space, $\mathcal{A}$ is required to be (possibly up to phase) an intertwiner between the representations $u_G$ and $\mathcal{U}_G$. $e^{i \theta_G} = \{\,e^{i \theta_g}\,|\,g \in G \}$ is a unitary character of $G$, and using the fact that the collection of these characters is isomorphic to $\mathcal{H}^1(G, U(1))$, the particular choice of $e^{i \theta_G}$ ends up specifying the 0D component of our SPTO signature.

With the virtual representation $\mathcal{U}_G: (H_{v}^*)^{\otimes 4} \to (H_{v}^*)^{\otimes 4}$ in hand, we can calculate the remaining portions of the SPTO signature of our state $\ket{\psi}$. The 2D portion of this signature relates to whether or not we can decompose $\mathcal{U}_G$ into a tensor product of four unitaries on the four virtual subsystems in $(H_{v}^*)^{\otimes 4}$. If we cannot, such that $\mathcal{U}_G$ is necessarily an entangled representation, then our state $\ket{\psi}$ has nontrivial 2D SPTO. In such cases, there are several (somewhat involved) procedures for extracting a 3-cohomology class to classify the 2D SPTO phase, but since our current interest is in the case of trivial 2D SPTO, we won't discuss these here. The interested reader can consult \cite{chen2011two, levin2012braiding, zaletel2014detecting} for examples of methods for obtaining information about 2D SPTO.

Given trivial 2D SPTO, we can write $\mathcal{U}_G$ as a tensor product of four terms, which we will assume has the form $\mathcal{U}_G = U_G^{(x)} \otimes (U_G^{(x)})^* \otimes U_G^{(y)} \otimes (U_G^{(y)})^*$. These four terms correspond to, in order, the left, right, top, and bottom portions of our virtual representation, where $(U_G^{(x)})^*$ (resp. $(U_G^{(y)})^*$) represent the complex-conjugated versions of $U_G^{(x)}$ (resp. $U_G^{(y)}$). We refer to $U_G^{(x)}$ and $U_G^{(y)}$ as the horizontal and vertical components of our virtual representation, and these determine the 1D portion of our SPTO signature. In particular, whether or not our system has nontrivial 1D SPTO is equivalent to whether or not the horizontal/vertical components of our representation are nontrivial projective representations of $G$. More concretely, the product of two elements of $U_G^{(\mu)}$, $U_g^{(\mu)}$ and $U_h^{(\mu)}$ ($\mu$ standing for either $x$ or $y$), will generally only equal $U_{gh}^{(\mu)}$ up to a phase factor, such that $U_{g}^{(\mu)} U_{h}^{(\mu)} = \omega_2^{(\mu)}\!(g, h)\, U_{gh}^{(\mu)}$. Multiplication of elements of $U_G^{(\mu)}$ is associative, and this condition ends up forcing our phases $\omega_2^{(\mu)}\!(g, h)$ to be 2-cocycles. The cohomology classes of these horizontal and vertical cocycles, $[\omega_2^{(x)}]$ and $[\omega_2^{(y)}]$, then form the 1D components of $\Omega_2$, the SPTO signature of $\ket{\psi}$.

Let's use these techniques to determine the SPTO signature of the 2D cluster state. We can choose a PEPS representation for a single qubit site of the 2D cluster state as $\mathcal{A}_{C}^{(1 \timess 1)} = \sum_{i=0}^1 \ket{i} \otimes A_i$, with the $A_i \in (H_{v}^*)^{\otimes 4}$ given by

\begin{equation}
    \label{eq:cluster_peps}
    A_0 = \bra{+X, 0, +X, 0} \ ,\ A_0 = \bra{-X, 1, -X, 1}.
\end{equation}

\noindent $H_{v}$ is here a qubit space, and the ordering of our systems in \eqnref{eq:cluster_peps} is as $(H_{v}^{(left)} \otimes H_{v}^{(right)} \otimes H_{v}^{(top)} \otimes H_{v}^{(bottom)})^*$. We are interested in the SPTO signature of the 2D cluster state with respect to a $2\timess2$ unit cell, since the cluster state then has its maximal on-site symmetry group of $G = \ztwofour$. To determine this, we contract together four copies of the PEPS tensor of \eqnref{eq:cluster_peps} to form a $2\timess2$ PEPS tensor, $\mathcal{A}_C^{(2 \timess 2)}$, and then find the virtual symmetry representations $U_G^{(x)}$ and $U_G^{(y)}$. These each act on a two-qubit virtual space, which for $U_G^{(x)}$ is decomposed as $(H_{v}^{(top)} \otimes H_{v}^{(bottom)})^*$, and for $U_G^{(y)}$ is decomposed as $(H_{v}^{(left)} \otimes H_{v}^{(right)})^*$.

As in the main text, we label the generators of $\ztwofour$ by their respective locations in the $2\timess2$ unit cell. One can then verify that the following choice of virtual symmetry representation makes our PEPS tensor $\mathcal{A}_C^{(2 \timess 2)}$ an intertwiner with respect to the physical representation $u_G$ (see Figure~\ref{fig:peps}c):

\begin{equation}
    \label{eq:virtual_representation}
    \begin{aligned}
    &U_{NW}^{(x)} = Z \otimes I  &   &U_{NW}^{(y)} = Z \otimes I  \\
    &U_{NE}^{(x)} = X \otimes I  &   &U_{NE}^{(y)} = I \otimes Z  \\
    &U_{SE}^{(x)} = I \otimes X  &   &U_{SE}^{(y)} = I \otimes X  \\
    &U_{SW}^{(x)} = I \otimes Z  &   &U_{SW}^{(y)} = X \otimes I
    \end{aligned}
\end{equation}

\noindent The fact that we can choose a form for $\mathcal{U}_G$ which factorizes into parts and satisfies \eqnref{eq:virtual_symmetry} with $e^{i \theta_G} = 1$ is confirmation of the trivial 2D and 0D SPTO of the 2D cluster state. The only thing that remains is determining the two 1D components of the SPTO signature. We can show that these are both nontrivial by considering the commutation relation of elements of $U_{G}^{(x)}$ and $U_{G}^{(y)}$. While $\ztwofour$ is abelian, the virtual representations in \eqnref{eq:virtual_representation} aren't, as shown by $U_{NW}^{(x)} U_{NE}^{(x)} (U_{NW}^{(x)})^\dagger (U_{NE}^{(x)})^\dagger = U_{NW}^{(y)} U_{SW}^{(y)} (U_{NW}^{(y)})^\dagger (U_{SW}^{(y)})^\dagger = -I^{\otimes 2}$. This means that the 2-cocycle $\omega_2^{(\mu)}$ associated with each of our virtual representations is different from the identity. Furthermore, multiplying either of these 2-cocycles by an arbitrary 2-coboundary is equivalent to modifying the phases associated to our individual $U_g^{(\mu)}$ as $U_g^{(\mu)} \mapsto \omega_1(g) U_g^{(\mu)}$, with $\omega_1(g) \in \mathcal{C}^1(G, U(1))$. This has no effect on the commutators of our symmetry group, which proves that our 2-cocycles $\omega_2^{(x)}$ and $\omega_2^{(y)}$ are in nontrivial 2-cohomology classes. The SPTO signature of the 2D cluster state is therefore $\Omega_2^{(C)} = \sptsignature{0}{1}{1}{0}$, meaning trivial 2D SPTO and nontrivial 1D SPTO, with the latter belonging to the nontrivial $D_2$ Haldane phase.

\section{Proof of the Pauli Universality of Our Resource State}
\label{sec:protocol_details}

In this Section, we give a proof of the fact that our Union Jack resource state is Pauli universal, meaning that it can carry out universal MQC using only measurements of single-qubit Pauli operators. Achieving this universality requires several components, namely:

\begin{itemize}
\item We can convert regions of our Union Jack to ``cluster regions", which are locally isomorphic to the 2D cluster state. This involves carrying out a pattern of computational basis measurements which converts (a part of) our state to a random graph state. The protocol of \cite{browne2008phase} (which uses only Pauli measurements) is then used to concentrate this state into a 2D cluster state, which in turn requires the percolation problem associated with our random graph states to lie in a supercritical phase. We demonstrate the supercriticality of this percolation problem, and thereby the ability to prepare cluster regions within our state, in Section~\ref{sec:conversion_to_cluster}.

\item We can teleport states and implement Clifford operations on them within the cluster regions of our state, using only Pauli measurements. Due to these cluster regions being identical to connected regions of the cluster state, we can use the same measurement patterns described in \cite{raussendorf2003measurement} to implement these Clifford operations, which use only Pauli measurements.

\item We can create ``interaction gadgets", which implement a three-qubit non-Clifford operation, $U_I^{(123)} = CCZ^{(123)} \sqrt{CZ}^{(12)} \sqrt{CZ}^{(23)}$, using only Pauli-basis measurements. Furthermore, these gadgets can be connected to a surrounding cluster region with a finite success probability, allowing us to use these gadgets as logical gates which we can connect together to create a $CCZ$ operation. We demonstrate these various facts in Section~\ref{sec:interaction_gadget}.

\end{itemize}

\noindent Taken together, these various facts successfully demonstrate the Pauli universality of our Union Jack state.

\subsection{Conversion to a 2D Cluster State}
\label{sec:conversion_to_cluster}

\begin{figure}[t]
  \centering
  \hbox{\hspace{1.5cm}
  \includegraphics[width=0.28\textwidth]{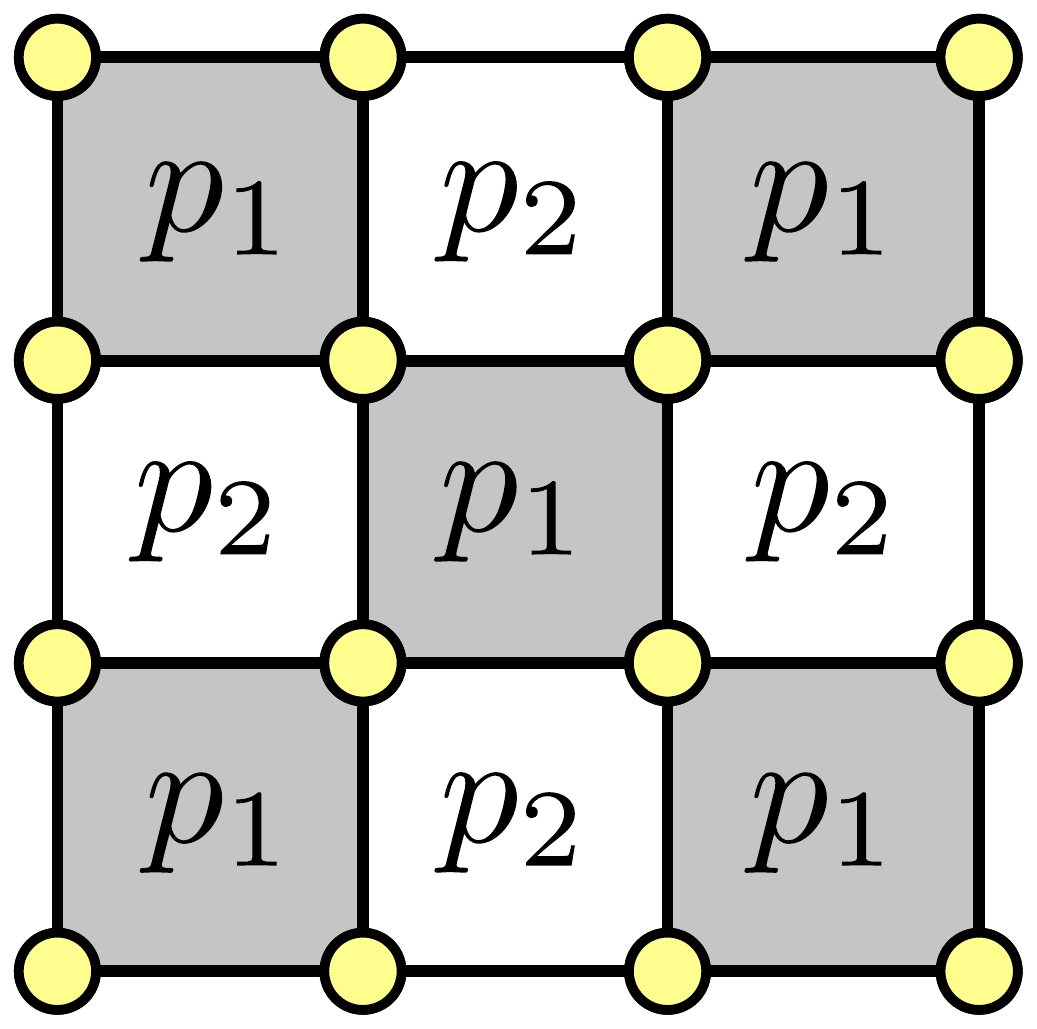}}
  \caption{A layout of our two-parameter percolation model. Cells labeled with $p_i$ ($i = 1,2$) are independently sampled, such that the probability of obtaining an outcome of 1 in that cell is $p_i$. An edge of our random graph state is set when two adjacent nodes differ in their values. This yields a deterministically empty lattice at $(p_1,p_2) = (0,0)$ or $(1,1)$, and a deterministically full lattice at $(p_1,p_2) = (0,1)$ or $(1,0)$. Additionally, setting $p_1 = 0$ (resp. $p_2 = 0$) gives a percolation problem which is isomorphic to a site percolation problem on a square lattice with a bond probability of $p_2$ ($p_1$). Our problem of interest is located at $(p_1,p_2) = (\onehalf,\onehalf)$.}
  \label{fig:checkerboard}
\end{figure}

\begin{figure}[t]
  \centering
  \hbox{\hspace{-0.25cm}
  \includegraphics[width=0.5\textwidth]{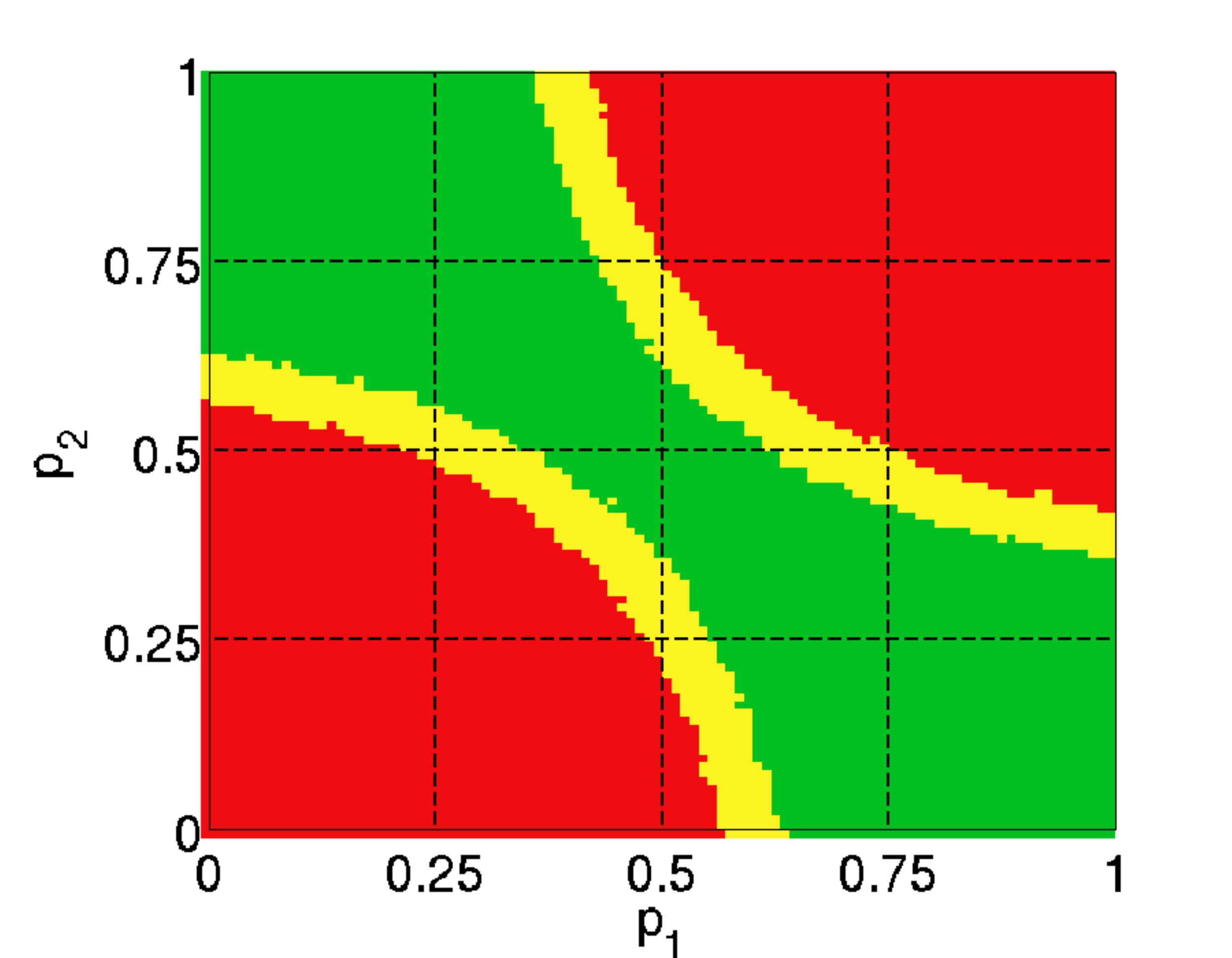}}
  \caption{The percolation phase diagram of our two-parameter model. Red (bottom left and upper right) indicates a subcritical phase, while green (upper left to bottom right) indicates a supercritical phase. The yellow region contains the critical line separating the phases. This division is based on the spanning probability $p_{span}$ when $m = 100$, and in particular whether $p_{span} \leq 0.05$, $p_{span} \geq 0.95$, or $0.05 < p_{span} < 0.95$. From the placement of our problem of interest at $(p_1,p_2) = (1/2,1/2)$, it is clear that we are within a supercritical phase, and can therefore use our 2D SPTO state as a universal resource for MQC.}
  \label{fig:phase_diagram}
\end{figure}

\begin{figure}[t]
  \centering
  \hbox{\hspace{0cm}
  \includegraphics[width=0.5\textwidth]{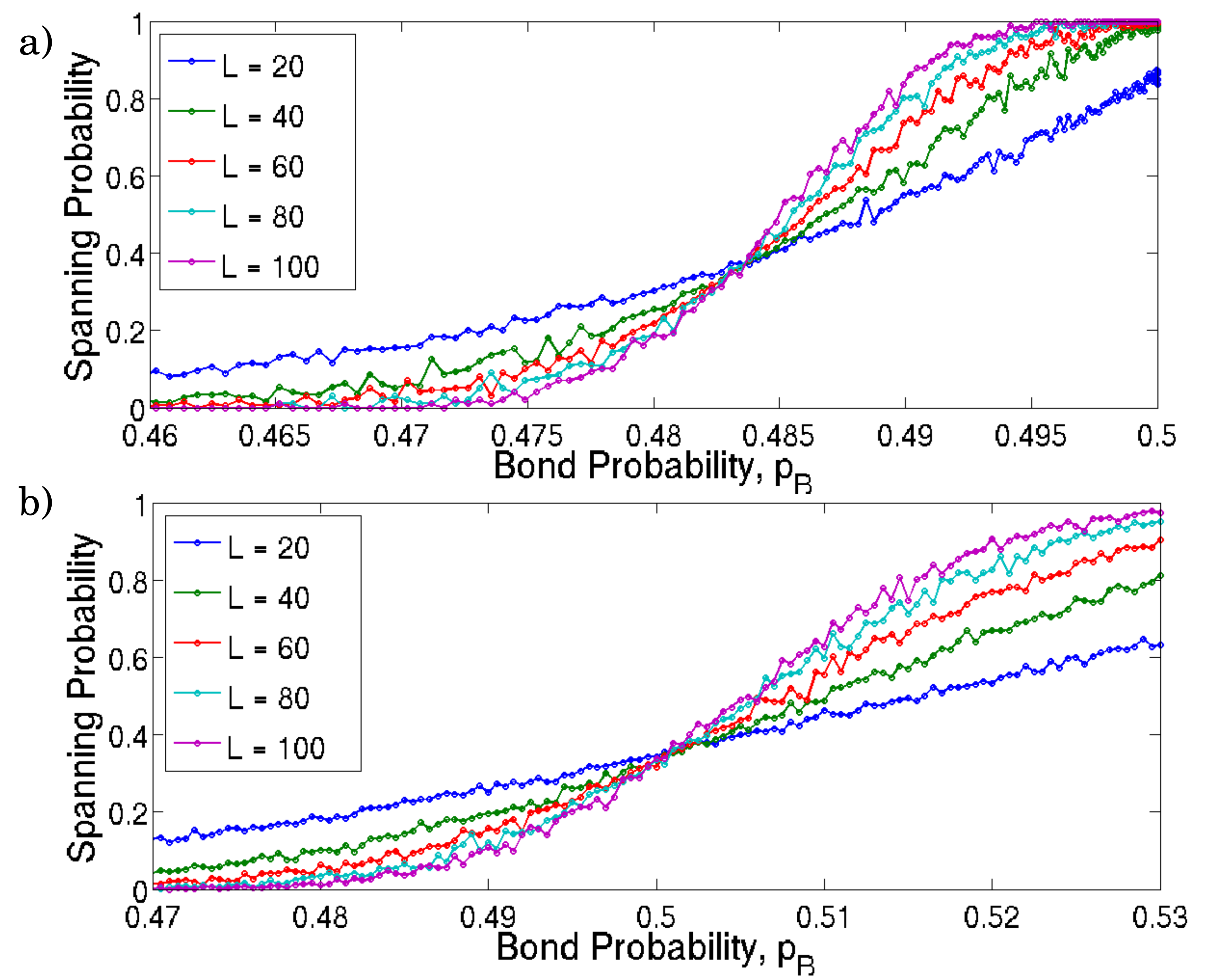}}
  \caption{a) The percolation probability for lattices of increasing linear size $L$, as we vary a parameter $\epsilon$ from 0 to 1. The marginal bond probability varies as $p_B = \epsilon (1 - \onehalf \epsilon)$, and the critical bond probability is seen to be $p_B = 0.484 \pm 0.001$. b) Using the same tools as were used in (a) to study the canonical square lattice bond percolation problem. The critical bond probability is known to be $\onehalf$, and our simulation reproduces this, locating it at $p_B = 0.500 \pm 0.001$.}
  \label{fig:phase_transition}
\end{figure}

After giving a more complete description of the reduction of our $\ztwo$ resource state to a random graph state, we describe the simulations we use to verify that the associated percolation problem is indeed in the supercritical phase. These simulations involve the construction of a two-parameter model which includes as a special case the percolation problem associated to our random graph state reduction protocol. We show that our particular percolation problem lies within a supercritical phase, thus demonstrating that the protocol of \cite{browne2008phase} can be used to efficiently convert these random graph states to a 2D cluster state with arbitrarily high probability.

As described in the Methods, the method we use for reducing our 2D SPTO resource state to a random graph state consists simply of measuring all of the control sites in the computational basis. Given $n$ control sites initially, upon measurement we obtain a string of random outcomes $\mathbf{c} = (c_1, c_2, \ldots, c_n)$. What is the reduced state of the logical portion of our system given a particular string of outcomes $\mathbf{c}$? To figure this out, we exploit the fact that the projector associated with our measurement outcome commutes with all of the $CCZ$'s, since the latter are diagonal in the computational basis. Thus, the state of our system after measurement is the same as if we had initialized the control sites in their post-measurement states, and afterwards applied $CCZ$ everywhere in our lattice. The resulting (unnormalized) state is then

\begin{equation}
    \label{eq:reduced_state}
    \ket{\tilde{\psi}(\mathbf{c})} = \frac{1}{\sqrt{2^n}} \prod_{ \ell \in \mathcal{L}_2} (CZ_\ell)^{c(\ell) + c'(\ell)} \ket{+X}^{\otimes n}.
\end{equation}

\noindent Here, $\mathcal{L}_2$ is the collection of edges in our lattice, $CZ_\ell$ is a controlled-Z gate applied to the endpoints of a logical edge $\ell$, while $c(\ell)$ and $c'(\ell)$ are the measurement outcomes obtained on the two control sites adjacent to $\ell$. The factor of $1 / \sqrt{2^n}$ emerges from the inner product of our $n$ measurement outcomes $\bra{0}$ or $\bra{1}$ with the $\ket{+X}$'s which were used to initialize our state. What \eqnref{eq:reduced_state} tells us (ignoring normalization) is that whenever the measurement outcomes on two adjacent control sites are not equal, a $CZ$ operation is performed on the logical edge in between them, while nothing is done when the measurement outcomes are the same.

From this description, it is easy to see that every state $\ket{\psi(\mathbf{c})}$ is a graph state, whose edges lie only along domain walls of the control site measurement outcomes. The control site outcomes themselves are uncorrelated and uniformly distributed, which follows from the equal magnitude of all of the unnormalized reduced states in \eqnref{eq:reduced_state}. More precisely, the probability of obtaining a particular outcome $\mathbf{c}$, $p(\mathbf{c})$, is given by

\begin{equation}
    \label{eq:measurement_probs}
    p(\mathbf{c}) = \braket{\tilde{\psi}(\mathbf{c})}{\tilde{\psi}(\mathbf{c})} = \frac{1}{2^n}.
\end{equation}

\noindent Ignoring the quantum origin of the probabilities, this probabilistic reduction to a graph state can be seen as defining a (classical) percolation problem, wherein edges of a graph are filled based on the configuration of random control site variables. We wish to conclusively determine whether this percolation problem, with site probabilities given by \eqnref{eq:measurement_probs}, corresponds to subcritical or supercritical behavior in the large-system limit. More explicitly, from the known behavior of percolation problems, we expect that the probability of obtaining a connected graph component which connects arbitrarily distant portions of our lattice goes to either 0 or 1 as we make our system size larger, and we would like to know which of these possibilities holds.

To do this, we carry out numerical simulations of a two-parameter percolation model identical to ours, but with tunable probabilities for different control site outcomes. While \eqnref{eq:measurement_probs} corresponds to a probability of $\onehalf$ of obtaining 1 on any arbitrary control site, our variable model has probabilities of $p_1$ on one half of the sites, and $p_2$ on the other half of the sites. Figure~\ref{fig:checkerboard} shows the checkerboard-style layout of these sites. The percolation problem defined by our actual system then corresponds to the point $p_1 = p_2 = 1/2$.

Figure~\ref{fig:phase_diagram} shows a phase diagram of this two-parameter model which demarcates the approximate locations of the subcritical and supercritical percolation phases. Although we haven't attempted to determine the exact location of the line of criticality which separates these two phases, it is clear that our system lies within the supercritical percolation phase.

Figure~\ref{fig:phase_transition}a shows the spanning probability we obtain along a one-parameter path through our configuration space. The path, parameterized by $\epsilon$, travels along $p_1 = p_2 = \onehalf \epsilon$ for $0 \leq \epsilon \leq 1$. The marginal probability of obtaining a single bond in our lattice is $p_B = p_1 + p_2 - 2 p_1 p_2 = \epsilon(1 - \onehalf \epsilon)$ along our path. A percolation phase transition is seen to occur at $p_B = 0.484 \pm 0.001$. For comparison, in Figure~\ref{fig:phase_transition}b we show a simulation of the standard square lattice bond percolation problem, wherein bonds appear independently of each other with probability $p_B$. Using identical methods, we identify a phase transition at $p_B = 0.500 \pm 0.001$, in agreement with the known exact value of $p_B = \onehalf$.

These results, along with the percolation results in Figure~4, conclusively demonstrate the supercritical behavior of the random graph states obtained in our state reduction protocol, thus proving our ability to prepare cluster regions within our Union Jack state using only Pauli basis measurements.

\begin{figure}[t]
  \centering
  \includegraphics[width=0.5\textwidth]{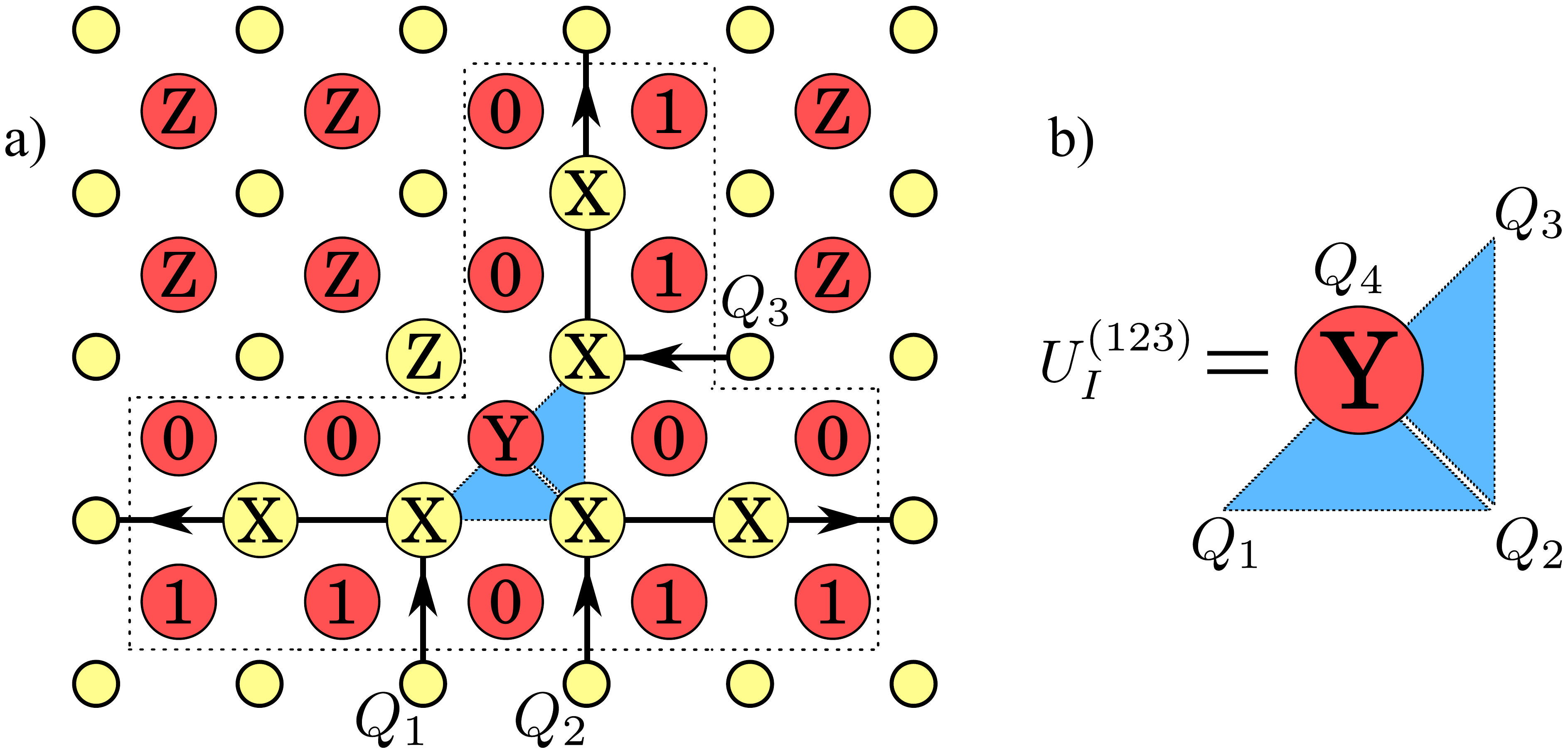}
  \caption{a) Our interaction gadget, which allows us to apply the gate $U_I$ to logical information. Blue triangles here represent $CCZ$ gates involved in forming the Union Jack state which play nontrivial roles in preparing $U_I$. We measure one control site in the Y-basis, six logical sites in the X-basis, and one logical site (along with many surrounding control sites) in the Z-basis, then use postselection to fix 13 of the control site measurement outcomes. The postselection is necessary to guarantee we can teleport information through the interaction gadget, the teleportation being carried out with the six X-basis measurements. b) The three-body operation which produces the diagonal unitary gate, $U_I$. Qubit 4 is initialized in a $\ket{+X}$ state, then contracted with a $\bra{+Y}$ outcome.}
  \label{fig:interaction_gadget}
\end{figure}

\subsection{Non-Clifford Gates using our Interaction Gadget}
\label{sec:interaction_gadget}

\begin{figure}[t]
  \centering
  \includegraphics[width=0.4\textwidth]{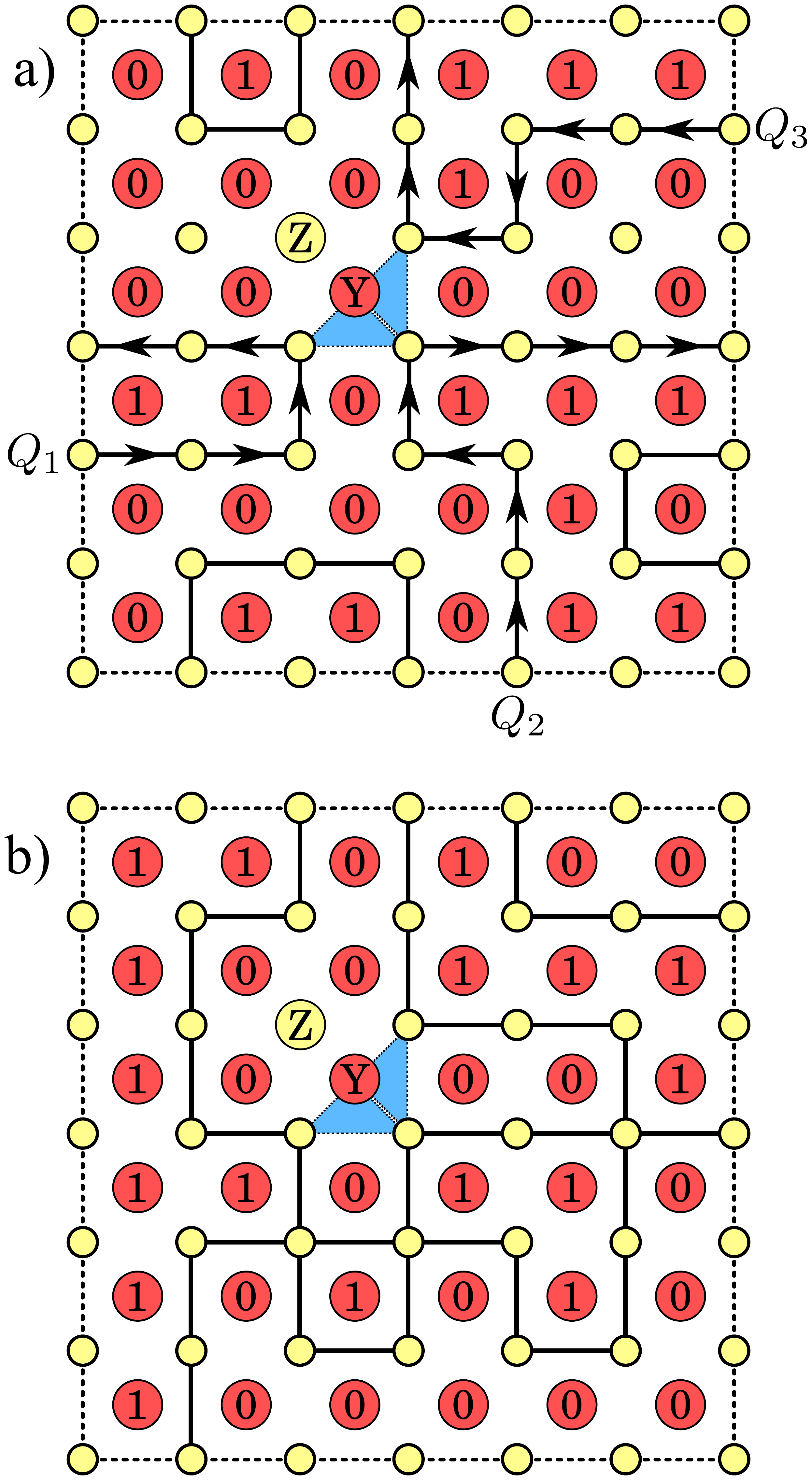}
  \caption{a) A pattern of control site outcomes which possesses the ``correct wiring" for our interaction gadget (X-basis measurements not shown). The wires percolate towards separate points on the boundary without intersecting each other and without being acted on by stray $CZ$ gates. b) An incorrect wiring pattern, which would require us to try again somewhere else in order to obtain a usable interaction gadget. Note that such regions can still be used as portions of cluster regions, without impacting the overall efficiency of our protocol. In this case the control site marked $Y$ would instead be measured in the computational basis.}
  \label{fig:interaction_gadget_wiring}
\end{figure}

We first prove that our interaction gadget, associated with the measurement pattern shown in Figure~\ref{fig:interaction_gadget}a, implements the unitary gate $U_I^{(123)} = CCZ^{(123)} \sqrt{CZ}^{(12)} \sqrt{CZ}^{(23)}$, and we give the Clifford byproduct operators associated with certain unintended measurement outcomes. We then discuss how such gadgets can be embedded into a surrounding cluster region, allowing them to act on arbitrary triples of qubits within that region.

The core of our interaction gadget is the three-body operation given by multiplying two overlapping copies of $CCZ$ and contracting one of the overlapping sites with an ancilla state $\ket{+X}$ and a Y-basis measurement outcome $\bra{\pm Y} = \tfrac{1}{\sqrt{2}}(\bra{0} \mp i \bra{1})$. Choosing \bra{+Y} to be the ideal outcome, this yields the operation

\begin{equation}
\label{eq:U_I}
U_I^{(123)} = \bra{+Y}^{(4)} \left( CCZ^{(124)} CCZ^{(234)} \right) \ket{+X}^{(4)},
\end{equation}

\noindent which is diagonal in the computational basis (shown in Figure~\ref{fig:interaction_gadget}b). Up to overall normalization and phase, $U_I^{(123)}$ gives a phase factor of $i$ when acting on $\ket{110}^{(123)}$ or $\ket{011}^{(123)}$, and a phase factor of 1 otherwise, proving that its operation is identical to $CCZ^{(123)} \sqrt{CZ}^{(12)} \sqrt{CZ}^{(23)}$. Because $\bra{-Y} = (\bra{+Y})^*$, the operation given by the outcome $\bra{-Y}$ is $(U_I^{(123)})^*$, which is equal to $U_I^{(123)}$ up to Clifford byproduct operators $CZ^{(12)} CZ^{(23)}$.

The three-body operation discussed above assumes that a $\bra{0}$ outcome has been obtained in the logical site $Z$ measurement adjoining the control site $Y$ measurement (yellow $Z$ in Figure~\ref{fig:interaction_gadget}a), and thus needs to be modified when a $\bra{1}$ outcome is obtained. In this latter case, the overlapping $CCZ^{(124)} CCZ^{(234)}$ in \eqnref{eq:U_I} is replaced by $CCZ^{(124)} CCZ^{(234)} CZ^{(14)} CZ^{(34)}$, and it can be shown that the resultant gate is again equal to $U_I^{(123)}$ up to Clifford byproduct operators $S^{(1)} S^{(3)}$, where $S$ is the phase gate $S = \mathrm{diag}(1, i)$. Finally, the case of unintended $Y$ and $Z$ outcomes in conjunction leads to Clifford byproduct operators $(CZ^{(12)} CZ^{(23)} S^{(1)} S^{(3)})^\dagger$.

In summary, we have shown that a combined $Y$ and $Z$ measurement is capable of converting two non-Clifford $CCZ$ gates into a three-body non-Clifford $U_I$ gate, with variation in measurement outcomes being accounted for by Clifford byproduct operators. Now how do we use this three-body unitary as a logical operation? One method for doing this is by measuring the control sites surrounding our interaction gadget in the computational basis, and then attempting to use the random graph state we obtain to teleport qubits through the sites which $U_I$ acts on. In the process of teleporting this information, $U_I$ is successfully applied to the three qubits of interest. However, we aren't guaranteed to obtain a graph state with the ``correct wiring", i.e. one for which we can separately teleport each logical qubit to and from its respective site adjoining the interaction gadget, as in Figure~\ref{fig:interaction_gadget_wiring}a. Because of the possibility of obtaining graph states with incorrect wiring patterns, the successful embedding of an interaction gadget into a surrounding cluster region only occurs with some probability, which generically depends on the size of the surrounding cluster region. 

We can show that the probability of obtaining a correct wiring pattern in the large system limit is finite and non-zero, by exploiting the same supercritical percolation properties which allowed us to prove the successful preparation cluster regions. This constant success probability then guarantees that the stochastic nature of our interaction gadget embedding only contributes a constant multiplicative factor to the number of sites measured in our protocol. Consequently, our MQC protocol gives a proof of principle demonstration that we can efficiently perform quantum computation. Our proof involves first restricting ourselves to a region of finite size surrounding a particular interaction gadget, then using postselection (with finite success probability) to obtain a pattern of control qubit measurement outcomes which prepares a graph state with the correct wiring. For example, choosing a $6 \times 6$ grid of control qubits, we could postselect for the pattern shown in Figure~\ref{fig:interaction_gadget_wiring}a.

When our region is of sufficient size, our postselected pattern can always be chosen so that distinct logical wires percolate without intersecting each other, and end at sufficiently separated points on the boundary of this region. When the separation between adjacent wire endpoints on the boundary of our finite region is much greater than the characteristic percolation length scale (the length scale associated with the exponential decay seen in Figure~4), the conditional probability of continuing our postselected pattern to a macroscopic graph state with the correct wiring factorizes into six uncorrelated probabilities. These probabilities, one for each wire, encode the chance of each wire percolating to a point infinitely far from its starting point on the finite region boundary. Because of the supercritical nature of this percolation, each of these conditional success probabilities is finite, meaning that the total success probability for embedding an interaction gadget in a large cluster region is finite. Thus, our interaction gadgets can be embedded in cluster regions with a constant multiplicative overhead, letting us efficiently use them as logical gates which, together with the Clifford gates we get from our cluster regions, form a universal gate set.


\begin{thebibliography}{99}
\bibitem{footnote1}
    We don't discuss SPTO of the 2D cluster state with respect to time reversal, inversion, or lattice rotations, as these symmetries wouldn't alter Theorem~\ref{thm:cluster_state}.
\end{thebibliography}
\end{document}